\documentstyle[11pt,aaspp4,epsf,epsfig]{article}

\def\beq{\begin{equation}}
\def\eeq{\end{equation}}
\def\beqar{\begin{eqnarray}}
\def\eeqar{\end{eqnarray}}

\def\he#1{\hbox{${}^{#1}$He}}
\def\li#1{\hbox{${}^{#1}$Li}}
\def\yp{\hbox{$Y_{\rm p}$}}
\def\msol{\hbox{${M}_{\odot}$}}
\def\outf{\vartheta}

\def\hii{{\rm H} {\footnotesize II}}
\def\etal{{\it et al.}~}
\def\minf{m_{\rm inf}}
\def\msup{m_{\rm sup}}
\def\effSN{\epsilon_{\rm SN}}
\def\effISM{\epsilon_{\rm ISM}}
\def\xhbb{\xi_{\rm hbb}}

\def\pcite#1{(\cite{#1})}
\def\pref#1{(\ref{#1})}

\def\bline{\rule[1.2mm]{3em}{0.1mm}}

\def\la{\mathrel{\mathpalette\fun <}}
\def\ga{\mathrel{\mathpalette\fun >}}
\def\fun#1#2{\lower3.6pt\vbox{\baselineskip0pt\lineskip.9pt
  \ialign{$\mathsurround=0pt#1\hfil##\hfil$\crcr#2\crcr\sim\crcr}}}

\begin{document}
\rightline{UMN-TH-1625/98}
\rightline{astro-ph/9803297}
\rightline{March 1998}

\title{ON THE EVOLUTION OF HELIUM IN BLUE COMPACT GALAXIES}

\author{Brian D. Fields and Keith A. Olive}
\affil{School of Physics and Astronomy, University of Minnesota \\ 
Minneapolis, MN 55455, USA \\
{\tt fields@mnhepw.hep.umn.edu 
\hphantom{\&}
olive@mnhep.hep.umn.edu}}

\begin{abstract} 
We discuss the chemical evolution of dwarf irregular
and blue compact galaxies in light of recent data, new
stellar yields and chemical evolution models.
We examine the abundance data for evidence of \hii\ region 
self-enrichment effects, which would lead to correlations 
in the scatter of helium, nitrogen, and oxygen abundances 
around their mean trends.  The observed helium abundance trends
show no such correlations, though the nitrogen--oxygen trend does
show strong evidence for real scatter beyond observational error.
We construct simple models for the chemical evolution of these galaxies,
using the most recent yields of \he4, C, N and O in intermediate- and
high-mass stars. The effects of galactic outflows, which can arise both
{}from bulk heating and evaporation of the ISM, and from the partial escape
of enriched supernova ejecta are included. In agreement with other
studies, we find that supernova-enriched outflows  can roughly reproduce
the observed He, C, N, and O trends; however, in models that fit N versus
O, the slopes $\Delta Y/\Delta$O and $\Delta Y/\Delta$N consistently fall
more than $2\sigma$ below the fit to observations.  
We discuss the role of the models and their uncertainties in the
extrapolation of primordial helium from the data. We also explore the
model dependence arising nucleosynthesis uncertainties associated with
nitrogen yields in intermediate mass stars, the fate of $8-11 \msol$
stars, and massive star winds.

\end{abstract}

\newpage

\section{Introduction}

Helium-4 plays a central role in big bang
nucleosynthesis.  The calculation of
the primordial helium abundance
is robust, being only weakly sensitive
to the cosmic baryon-to-photon ratio, $\eta$.
However, the helium abundance is quite sensitive to,
and provides a strong constraint on the physics of the early universe.
The primordial abundance of helium (with mass fraction \yp)
is best determined via observations of
\hii\ regions in the most metal-poor galaxies---dwarf irregulars
and  blue compact galaxies (hereafter, BCGs).
For these systems, the helium evolution is derived
empirically:  the data show
that the helium abundance increases with metallicity indicators
such as 
oxygen and nitrogen.
Following Peimbert \& Torres-Peimbert, \pcite{ptp},
the primordial abundance of helium is inferred from an
extrapolation of the observed trend to zero metallicity.

As emphasized by
Fields \& Olive\pcite{fo};
Fields, Kainulainen, Olive, \& Thomas \pcite{fkot};
and Olive, Steigman, \& Skillman (\cite{oss}; hereafter OSS97),
the empirical extrapolation
of primordial helium 
sidesteps any reliance on detailed results of
chemical evolution models, especially since the extrapolation to zero
metallicity from the data is small.  
Indeed, until the measurements of D/H in quasar absorption systems 
can be confirmed to represent a uniform primordial value, the \he4
abundance is crucial when used in conjunction with \li7 as a test of big
bang nucleosynthesis theory.  (This remains true so long as 
$\yp \la 0.24$; at high \yp, 
the insensitivity of \yp\ to $\eta$ makes \he4 a very
poor discriminator for primordial nucleosynthesis.)  
Thus, given the importance of
\he4 as observed in BCGs, 
it is clearly of interest to compare the predictions
of chemical evolution with 
the observations of these systems.
A less than successful comparison of theory with observation
could point to the subtleties in chemical
evolution modeling of even these
(apparently) simple systems.

Several groups have modeled the chemical evolution of BCGs 
(Matteucci \& Chiosi \cite{mc};
Matteucci \& Tosi \cite{mt};
Gilmore \& Wyse \cite{gw};
Pilyugin \cite{pil};
Clayton \& Tantelaki \cite{cp};
Marconi, Matteucci \& Tosi \cite{mmt};
Carigi, Col\'{\i}n, Peimbert, \& Sarimento \cite{ccps}).
The high star formation activity in BCGs
indicates that star formation must occur in stochastic bursts,
unless the systems are very young; all models
included this behavior.
In addition, these galaxies probably drive outflows of material,
perhaps with efficiencies high enough to significantly
alter the chemical and/or dynamical evolution.
For example, the gas fraction--metallicity relations of BCGs
may not be compatible with closed box models, suggesting
the importance of gas outflow via supernova-driven outflows
(Lequeux, Rayo, Serrano, Peimbert, \& Torres-Peimbert \cite{leq};
Matteucci \& Chiosi \cite{mc};
Carigi, Col\'{\i}n, Peimbert, \& Sarimento \cite{ccps})---though 
the difficulties of
determining an accurate gas--to--baryon fraction lead to significant
uncertainties in these arguments.  
In any case, X-ray observations of diffuse, hot gas
clearly support the existence
of outflows in some BCGs
(Heckman et al.\ \cite{heck}; Della Ceca, Griffiths, Heckman, 
\& Mackenty \cite{dc_etal}; 
Della Ceca, Griffiths,  \& Heckman \cite{dcgh}),
though Skillman \pcite{evan96}
and Skillman \& Bender \pcite{sb} present
arguments against a dominant role of
outflows in {\em all} dwarf galaxies.  
Independent of the question of outflows,
another key result from these studies 
showed that the usual instantaneous {\em mixing} approximation 
is inappropriate.
Self-enrichment of \hii\ regions in a burst phase 
(Kunth \& Sargent \cite{ksgt}; 
Kunth, Lequeux, Sargent, \& Viallefond \cite{klsv};
but see Pettini \& Lipman \cite{pl})
can lead to significant scatter in abundance trends.

Most recently, it has been argued that the due to the intense bursts of
star formation and energy release in the evolution of these systems, 
their stability requires of the presence of significant amounts of dark
matter (Bradamante, Matteucci, \& D'Ercole \cite{bmd}).  This may have an
impact on the effect of supernovae driven winds on the elements
abundances.
In a somewhat different approach,
Mac Low \& Ferrara \pcite{mlf} take the presence of dark
matter as a starting point, and examine the 
requirements for
BCG outflows as a function of supernova rate and galaxy mass.
Their models of the
the dark matter potential, and the detailed gas dynamics of 
outflows, suggest that the loss of the entire ISM---``blowaway''---is 
only possible for smaller galaxies ($\la 10^6 \msol$).
However, Mac Low \& Ferrara also find that the preferential 
escape of the energetic and metal-enriched supernova ejecta
is much easier, occurring at significant levels for 
galaxies with baryonic masses up to $\sim 10^8 \msol$.

While there has been great progress in laying out the basic
features of BCG chemical evolution, some important
issues remain unresolved, 
including some bearing on derivation of
primordial helium.
For example, there remains some question as to the 
theoretical justification for the linear fit of helium versus metals,
particularly nitrogen 
(e.g., Mathews, Boyd, \& Fuller \cite{mbf};
Balbes, Boyd, \& Mathews \cite{bbm}).
Closely related to this is 
the dependence of the inferred \yp ~on 
the metallicity tracer.  
Several studies (e.g., OSS97; Olive, Steigman, \& Walker \cite{osw})
 have found that 
the primordial helium abundance derived from a
linear fit to $Y-$N 
consistently differs  by
about $1\sigma$
{}from that derived via $Y-$O.
While this difference is as yet too mild to be a problem, 
it does suggest need to re-examine the appropriateness
of the linear fits in the context of detailed chemical
evolution models.
Recent new observations such as the growing set of carbon observations
are ripe for more theoretical attention.

In this paper, we examine the chemical evolution
of BCGs, with particular attention to these
issues related to primordial helium,  
and the uncertainties in the chemical evolution modeling.
In particular, we note the interplay between galactic
outflow prescriptions and nucleosynthesis uncertainties.
Different parameterizations for the nucleosynthesis of
nitrogen and their role \yp(N) versus \yp(O) discrepancy
are also considered.
We note the model-dependence of $\Delta Y/\Delta Z$ and other slopes,
and examine in detail a suggestion by Fields \pcite{fields}
that helium production processes due to low-mass supernovae
or high-mass stellar winds may account for the large observed
values of the slopes.  
We conclude that the empirical fitting procedures used
to obtain \yp\ do find support from our models.
However, we also note that models using the most recent
and detailed nucleosynthesis yields do no reproduce the
observed $Y-$N,O trends, even in the presence of outflows.
Possible solutions to this problem are discussed.

\section{Data}
\label{sect:data}

The present observations  of helium and metal tracer (N and O) 
abundances in BCGs 
are summarized in OSS97.
Their analysis included 
the data of Pagel \etal \pcite{p},
Skillman \& Kennicutt \pcite{sk}, Skillman \etal \pcite{setal}, Skillman
(private communication), and  
Izotov, Thuan, \& Lipovetsky (\cite{itl1}; \cite{itl2}, hereafter ITL97).
Though the ITL97
set was shown to be consistent with the prior data on a point by point
basis, a linear extrapolation of the ITL97 set alone
yields a significantly different slope in the $Y-$O/H plane,
and a higher primordial \he4\ abundance.
This suggests a systematic discrepancy, a
problem further accentuated in the more recent data of Izotov \& Thuan
(\cite{it2}, hereafter, IT), which we comment on below.

One other metal tracer observed in BCGs is carbon
(Garnett et al.\ \cite{gsds};
Kobulnicky \& Skillman \cite{ks};
Garnett et al.\ \cite{donetal}; ),
Torres-Peimbert, Peimbert, \& Fierro \pcite{tppf}).   
C/H data is available for far fewer systems than are N/H and O/H,
and so we will not rely on it as a 
metallicity tracer in extrapolating $\yp$.
However, the amount of carbon data is growing, and 
can provide a useful discriminant of 
different evolution schemes (Steigman, Gallagher, \& Schramm 
\cite{sgs}; Walker \etal \cite{wssok}, Olive \& Scully \cite{osc};
Carigi et al.\ \cite{ccps}).

\begin{table}[htb]
\caption{Observed Helium Fit Parameters}
\label{tab:Y_fit}
\vskip .1in
\begin{tabular}{|lcc|cr@{$\, \pm \,$}lc|}
\hline\hline
\multicolumn{7}{|c|}
   {$Y = \yp(A) + \frac{\Delta Y}{\Delta A} \, A$} \\
\hline
Data                  & No. of   & Tracer & & \multicolumn{2}{c}{} & \\
  set${}^{({\rm a})}$ & regions  &   $A$  &  
   $\yp(A)$  &  \multicolumn{2}{c}{$\Delta Y/\Delta A$}  &  $\chi^2_\nu$  \\
\hline\hline
B98 & 73 & N/H & $0.241 \pm 0.001$ & 1927&448 & 0.70 \\
    &    & O/H & $0.238 \pm 0.002$ & 112&25   & 0.66 \\
    &    & $Z$${}^{({\rm b})}$ & \verb+"+ & 5.0&1.1 & \verb+"+ \\ 
\hline\hline
C98 & 36 & N/H & $0.238 \pm 0.002$ & 3483&1587 & 0.80 \\
    &    & O/H & $0.235 \pm 0.003$ & 157&61    & 0.73 \\
    &    & $Z$ & \verb+"+ & 7.0&2.7 & \verb+"+ \\ 
\hline                        
IT (B) & 44 & N/H & $0.242 \pm 0.002$ & 1985&598 & 0.64 \\
       &    & O/H & $0.239 \pm 0.002$ & 108&32   & 0.63 \\
       &    & $Z$ & \verb+"+ & 4.8&1.4 & \verb+"+ \\ 
\hline                        
PS (B) & 40 & N/H & $0.236 \pm 0.003$ & 2311&725 & 0.53 \\
       &    & O/H & $0.234 \pm 0.003$ & 116&39   & 0.57 \\
       &    & $Z$ & \verb+"+ & 5.2&1.7 & \verb+"+ \\ 
\hline\hline
Carbon & 9 & C/H & 
   $0.232 \pm 0.004$ & 550&214 & 0.30 \\                        
\hline\hline
\end{tabular} \\
Notes:  ${}^{({\rm a})}$See text for description of data sets.\\
\hphantom{Notes:} ${}^{({\rm b})}$Metallicity assumed to
scale with oxygen:  $Z = ({\rm O/O}_\odot) \, Z_\odot = 22 \, {\rm O/H}$.
\end{table}

\begin{table}[htb]
\caption{Observed Nitrogen {\em vs.} Oxygen Fit Parameters}
\label{tab:NO_fit}
\begin{tabular}{|lc|cr@{$\, \pm \,$}lc|ccc|}
\hline\hline
Data & No. of  & 
  \multicolumn{4}{|c|}{${\rm N/H} = \alpha {\rm O/H} + \beta ({\rm O/H})^2$} & 
  \multicolumn{3}{c|}{${\rm N/H} = \zeta ({\rm O/H})^\omega$} \\
set  & regions & $\alpha$ & \multicolumn{2}{c}{$\beta$} & $\chi^2_\nu$ &
  $\log \zeta$ & $\omega$ & $\chi^2_\nu$ \\
\hline 
B98     & 73 & $0.022 \pm 0.002$ & 104&20 & 3.96 & 
      $-0.15 \pm 0.42$ & $1.32 \pm 0.05$ & 3.86\\
C98     & 36 & $0.018 \pm 0.002$ & 202&44 & 2.65 & 
      $-0.06 \pm 0.63$ & $1.34 \pm 0.08$ & 2.31\\
IT (B)  & 44 & $0.022 \pm 0.002$ & 100&22 & 5.59 & 
      $-0.51 \pm 0.41$ & $1.24 \pm 0.05$ & 4.56\\
PS (B)  & 40 & $0.025 \pm 0.003$ & 137&39 & 2.37 & 
      $-0.10 \pm 0.59$ & $1.32 \pm 0.07$ & 2.12\\
\hline
B98 & 73 & & \multicolumn{2}{c}{} & 
     & $-1.50 \pm 0.48$ & $\equiv 1$ & 435 \\
    &    & & \multicolumn{2}{c}{} & 
     & $2.64 \pm 0.81$ & $\equiv 2$ & 707 \\
\hline\hline
\end{tabular}
\end{table}

OSS97 divide the available data into sets A, B, and C, based on different
cuts in metallicity, and perform the bulk of their analysis on the
latter two. Set B contained 62 distinct \hii\ regions with 
O/H $< 1.5 \times 10^{-4}$ and N/H $< 1.0 \times 10^{-5}$,
omitting all but a few systems
{}from the total observed.
In view of the importance of points at very low metallicity,
OSS97 also construct set C,
which contained only the systems (32 in total),
for which ${\rm O/H} < 8.5 \times 10^{-5} \approx 10^{-1} {\rm
O/H}_\odot$.
Since the work of OSS97, IT have extended their data set which adds 12
points to set B (hereafter, B98) and 4 to set C (C98).
In our analysis, we will use this enlarged data set which contains the
bulk of the available data on the abundances of \he4, O, and N.
We have made one modification, however, to the IT data.  In some cases,
they present data with an unusually high degree of accuracy (less than
half a percent in some instances). While the quoted error may reflect
the uncertainty they considered, 
it was argued by Skillman \etal \pcite{setal} that 
abundance measurements can not be made to an accuracy of
better than 2\% at this time. For \he4, this corresponds to an
uncertainty of about 0.005 for an individual point. Therefore we have
systematically imposed a lower limit on the uncertainty of any abundance
measurement of 2\%.

OSS97 
examined the trends in the three axes of HeNO space,
namely $Y-$N, $Y-$O, and N$-$O.  
Following Peimbert \& Torres-Peimbert \pcite{ptp},
linear fits of $Y$ versus tracers N and O are
used to infer the intercepts \yp.  The parameters
of these fits appear in Table \ref{tab:Y_fit};
the data themselves appear in Fig.\ \ref{fig:NOcut}.
In Table \ref{tab:Y_fit}, the
updated HeNO data sets B98 and C98 are used to determine $\yp$ (the
intercept of the regression) and the slope with respect to both O/H and
N/H, and we have added linear fits to the carbon data (Olive
\& Scully \cite{osc}) from  9 \hii\ regions. 
Izotov \& Thuan \pcite{it1} have also argued 
that the I Zw 18 NW data
is unreliable due to the effects of underlying stellar absorption
and we omit this point from these data sets. 
Note that due to the relatively small errors quoted in IT, this data set
has come to dominate the entire sample.
For comparison, Table \ref{tab:Y_fit} also shows
the separate results for the ``B'' sets from IT and from the OSS97 set
minus IT, equivalent to the data treated in Olive \& Steigman \pcite{os}
without I Zw 18 NW (labeled PS in the table).
The higher values of \yp, quoted by IT are based on their
\he4 abundance derived by their method of determining all of the
parameters from a set of 5 helium recombination lines. This gives 
$\yp = 0.2444 \pm 0.0015 + (44 \pm 19)$O/H.  
However, as argued in OSS97 there are inherent
uncertainties in this method which are not reflected in the error budget.
For this reason and because we can more easily compare their data with
previous data, we use their results which are based on S {\footnotesize II} 
densities. 

In all of the data sets in Table \ref{tab:Y_fit},
the helium slopes $\Delta Y/\Delta A$ 
are several standard deviations away from
zero---the increase of helium versus metal tracers is  highly
statistically significant as was argued in Olive, Steigman \& Walker
\pcite{osw}.  Note
also that the $Y$ versus N and O fits have a
$\chi^2$ per degree of freedom
$\chi^2_\nu < 1$.  This suggests
that the scatter in $Y$-NO is entirely consistent
with the (statistical) observational uncertainties.
Indeed, the small $\chi^2_\nu$ suggest that the quoted
errors might be underestimated (though it does not address
the thorny issue of systematic errors).

We will take the B98 data as our set of
BCG abundances to model; the key parameter
in judging the He evolution we calculate will be
the slopes of the linear regressions.
For brevity, we will denote the ``helium slope''
$\Delta Y/\Delta {\rm O} \equiv \Delta Y/\Delta ({\rm O/H})$,
with a similar expression for N.
For comparison with previous work,
Table \ref{tab:Y_fit} also quotes values for 
$\Delta Y/\Delta Z$, with 
$Z \equiv Z_\odot ({\rm O/H})/({\rm O/H})_\odot = 22 {\rm O/H}$.
However, in this paper we will emphasize
$\Delta Y/\Delta$O and $\Delta Y/\Delta$N,
which are closer to the data and avoid 
questions of how to infer $Z$ from the observations.

OSS97 also examined the growth of nitrogen as a function of
oxygen, as a diagnostic of a component of ``secondary'' nitrogen.  
Such a component would arise if the stellar nitrogen yield is proportional
to the initial abundance of a ``seed'' isotope such as ${}^{12}$C.
Parameters for a quadratic fit, and a power law fit,
appear in Table \ref{tab:NO_fit}.  Both of these fits
are consistent with a combination of primary and secondary nitrogen,
(and worsen significantly if one demands that N is purely 
primary or secondary).  Furthermore, the $\chi^2_\nu \ga 3$
indicates that the scatter in the data is not consistent
with the (statistical) observational errors alone.
In other words, the scatter must be due to either systematic
errors, or an intrinsic dispersion in the data.  We will
examine this point further below (\S \ref{sect:scatter}).

Given the nonlinear dependence of nitrogen on oxygen, it follows that
the helium cannot increase linearly with both of these tracers.
Specifically, oxygen is almost certainly ``primary'' and so on general grounds
one can expect $Y-$O to be at least roughly linear.  However,
since nitrogen has a nonlinear, secondary component (versus O/H), 
one would expect that $Y-$N is not a linear trend.  
The departure from linearity
(as inferred from combining a linear $Y-$O and quadratic N--O)
is not large, and does not set in until the quadratic term
in N--O is roughly comparable to the linear term.
This transition is around 
${\rm O/H} \simeq \alpha/\beta = 212 \times 10^{-6} = 0.25 \, {\rm O/H}_\odot$,
or at 
${\rm N/H} \simeq 2 \alpha^2/\beta = 
93 \times 10^{-7} = 0.08 \, {\rm N/H}_\odot$.
Note that these regions coincide nicely with the
metallicity cuts that define data set B98.  
Still, in this metallicity range,
$Y$ versus N still displays some sensitivity to the
secondary component of N.  If the quadratic component of
the N--O fit were truly negligible (i.e., $\beta \simeq 0$),
then one would expect the slopes in the HeNO planes to be
related by $\Delta Y/\Delta O = \alpha  \Delta Y/\Delta N$;
in fact, this relation is off by $2\sigma$, due to the
effect of secondary N in depressing $\Delta Y/\Delta N$.

Recently, Lu, Sargent, \& Barlow \pcite{lsb}
have provided complementary observations relevant
to the N$-$O trends in low metallicity 
(down to [Si/H]$\sim -2$) systems.
They used the N/Si ratio in damped Lyman-$\alpha$ 
quasar absorption systems
as a diagnostic for N/O at high redshift,
under the well-motivated assumption that
silicon is predominantly produced by type II supernovae,
and thus traces oxygen.
The Lu, Sargent, \& Barlow \pcite{lsb}
data is best fit by a combination of
primary and secondary nitrogen, consistent with
the BCG observations.
Furthermore, the quasar absorber data
suggests that N shows increased scatter at low O/H.
These observations have considerable uncertainties due to, e.g., 
possible effects of N or Si depletion onto dust grains;
nevertheless Lu et al.\ argue that the basic trends
are real, and are consistent
with a ``time-delayed'' origin of nitrogen
in low-mass stars, 
which we investigate in the next section.

Finally, van Zee, Salzer, \& Haynes \pcite{vzsh} have recently
observed N, and O in low metallicity 
spiral galaxies.
They find no evidence of a real difference
between the N--O trends in dwarfs and spirals,
except that the abundances in dwarfs extend
to lower metallicities.
They identify the transition point
between the low-metallicity, primary behavior
of N--O to the high-metallicity, secondary behavior;
they locate this point---the ``knee''---at
$12 + \log ({\rm O/H})_{\rm knee} = 8.45$.  
This is consistent
with our estimate above, that the onset of
nonlinearity occurs around ${\rm O/H} \simeq \alpha/\beta$,
or $12 + \log ({\rm O/H})_{\rm knee} = 8.33$.

\section{Self-Enrichment, Scatter, and Correlations}
\label{sect:scatter}

Chemical evolution models of BCGs have considered not only the origin
of the mean abundance trends, but also sources of dispersion about the
mean.  Several authors (Garnett \cite{don}; Pilyugin \cite{pil};
Marconi, Matteucci, \& Tosi \cite{mmt}) have noted that the bursting
star formation in the \hii\ regions leads to an {\em intrinsic} component
of dispersion.  This comes about essentially due to incomplete,
non-instantaneous mixing of burst ejecta, coupled with different
evolutionary timescales for He, N, and O{}.  Specifically, consider an
initially well-mixed \hii\ region in a BCG, with an initial composition that
falls on the mean HeNO evolution.  Upon a burst
of star formation,  high mass stars will explode first (as supernovae),
and the death of low mass stars will be delayed relative to this.  In
this way, the initial mass function of the burst is effectively
scanned over time, from the high mass end to the low.  
Self-enrichment of the bursting (i.e., \hii) region 
is a transient phenomenon; it occurs in the time interval
when the burst's high-mass stars have begun to die and
inject their nucleosynthesis products, but before
a significant portion of the low-mass stars have done so.
Because the high-mass stars only produce some elements, 
their enrichment drives
the region's composition away from the mean (well-mixed) HeNO trends.
The bursting region only relaxes to the average 
instantaneous mixing trend after the low mass stars
finally return their ejecta.  Over a still longer timescale, the
enriched region would remix its composition with the rest of ISM.
This remixing may or may not occur before the next burst begins.

This scheme, therefore, predicts that intrinsic scatter inevitably
occurs at some level.
Furthermore, the self-enrichment mechanism predicts
{\em correlations} in the dispersion patterns.
Again, we are interested in a BCG with an (initially) well-mixed
composition falling on the mean trends in HeNO;
and we suppose there is a  burst of star formation
which forms an \hii\ region. 
Now consider the epoch of ``adolescence,'' 
when only the most massive stars
($m \ga 8 \msol \Rightarrow t \la 40$ Myr)
have died.
These stars make all of the burst's complement of new O, 
about half of the new He, and very little of the new N.
Placing this system on the space of HeNO abundances,
we see that compared to the mean trend, this system would show
a low N/H and a low $Y$ for its O/H abundance.  
Conversely, for its N/H, our adolescent system
would have a high $Y$ relative to the average.
These correlations should exist, and will be
observable in the data if the amplitude of the
intrinsic scatter is larger than the observational errors.

We now examine the data for evidence of these correlations.
To do this, we will divide the 3-dimensional HeNO data space
according to our 2-elements fits (which
suppress one element, e.g., fitting N versus O; 
see Tables \ref{tab:Y_fit} and \ref{tab:NO_fit}).  
These fits describe 2-dimensional surfaces in the data 
space, and we will divide the data into points
above and below the fit.  Then we will examine these
points in the two planes suppressed in the fit.
For example, using the OSS97 quadratic fit to
N$-$O (Table \ref{tab:NO_fit}), we cut the
N$-$O plane into regions of ``high N'' and ``low N''
(fig.\ \ref{fig:NOcut}a).  
In Figures \ref{fig:NOcut}b and \ref{fig:NOcut}c, 
we replot the high- and low-N points in the
$Y$--O/H and $Y$--N/H planes.

Let us assume for the moment that the scatter in each of
the three plots is real and due to self-enrichment.
Then, as argued in the preceding paragraph, the low N points represent
regions that are self-enriched; thus these points
should have {\em low} $Y$ relative to their
oxygen abundance and the mean $Y-$O trend.
Similarly, we expect the same points to have
{\em high} $Y$  relative to their
nitrogen abundance and the mean $Y-$N trend.
That is, there should be correlations
between a BCGs location on the N$-$O scatter and
its position on $Y-$NO; furthermore, the
expected clusterings in $Y-$N and $Y-$O are on opposite sides 
of the respective means.
However, these correlations do not appear in the data.
As seen in fig.\ \ref{fig:NOcut}b, both the
high and low N points fill the $Y-$NO spreads
more or less evenly.  The fraction of low N points
that are high in Figure \ref{fig:NOcut}b (i.e., above our $Y-$N fit) 
is $43 \pm 11\%$, compared to $37 \pm 7\%$
for high N points.\footnote{Here the errors are only $\sqrt N$
effects of counting statistics, implicitly assuming
perfect determination of the abundances---thus 
significantly underestimating the true error.}  
In the $Y-$O plane (Figure \ref{fig:NOcut}c), the 
the low N population
has a fraction $67 \pm 10\%$ of points with
high $Y$; the high N population has
 $58 \pm 7\%$.  Thus
the high and low populations in N--O show no significant difference
in the other two planes.
In particular, there is no
evidence that the low N points are biased high in $Y-$N
and low in $Y-$O.  

We can also perform a similar test for self-enrichment correlations
by splitting the data according to the location on
the $Y-$O plane.  As with the N--O split,
we are unable to find convincing evidence for the correlations
expected from self-enrichment.\footnote{There is a significant correlation
between the high $Y-$O and high $Y-$N populations,
but this is simply due to the shallow slopes of each trend---which
means that the He points are just ``reshuffled''---and to the
relative tightness of the N--O relation, which prevents
significant reshuffling.  This strong correlation is not, however,
the one expected from self-enrichment.}
One can understand these results if 
the dispersion in $Y-$N and $Y-$O is largely observational,
as indeed strongly suggested by the low $\chi^2_\nu < 1$ for
these data.  
The preceding discussion confirms this lack of real dispersion,
even if one were to assign the observations smaller errors.
This is not to say that intrinsic spread is absent, but rather
that considerably more accurate data will be needed to detect it.
However, the small $\chi^2_\nu$ does suggest that whatever scatter
is present must be relatively small--certainly less than the observational
errors.  

On the other hand, the N--O fit has $\chi^2_\nu = 3.3-3.4$;
thus it is quite likely that the N--O data
{\em does} contain scatter beyond the observational error.  
While some of this could be systematic, it is certainly
possible that some is real.  Thus, models to explain 
the scatter must account for the N--O spread
without introducing too much dispersion in
$Y$ versus N and O. Also, as discussed below, Pilyugin \pcite{pil} and
Marconi, Matteucci, \& Tosi \pcite{mmt}, 
predict that the N--O scatter amplitude is much larger than the $Y-$N,O
scatter.  Indeed, BCGs could well undergo enriched
outflows which preferentially remove massive star ejecta (see below);
this effect would suppress the degree of self-enrichment
in an adolescent \hii\ region.

In light of the real scatter in the BCG nitrogen versus oxygen
trends, we now turn to the 
Lu, Sargent, \& Barkow \pcite{lsb}  
observations of N/Si ($\propto$ N/O)
in damped Ly$\alpha$ absorbers (DLA).
The observations are at $z > 2$, which implies
$t \la 4$ Gyr for most reasonable cosmologies, which
is an early epoch, but still late enough
that stars with $m \ga 2 \msol$ can already have died.
Low-mass stars can thus have produced the nitrogen
in these systems.
In fact, Lu et al.\ find that the N/Si scatter increases
at low metallicity, and suggest that this is arises
due to the longer timescales for N production in 
low-mass stars.  

While N/Si in DLAs is related
to the N$-$O behavior in BCGs, the two
systems are likely very different.
DLAs are probably large galaxies
similar to own, and significantly bigger than 
BCGs.  A larger size of the DLAs would suggest
star formation proceeding in a more continuous manner,
where stochastic bursting effects would be smaller or nonexistent
Consequently, for an individual DLA, there would be less spread
due to bursting self-enrichment as in BCGs.  
However, if there is a variance in the ensemble of 
DLA star formation histories, 
one would predict a spread in N$-$O.  That is, if different
DLAs at fixed $z$ have different star forming timescales or amplitudes,
this would cause variations in not only O (which measures the
total integrated star formation), but also N/O
(which measures the fraction of star formation occurring
over low-mass star lifetimes).  
We note two scenarios in which an {\it increased} 
scatter in N$-$O at low metallicities is a  
measure of the variance in DLA star formation histories.
(1)  The early star formation in DLAs has significant variance
in starting time
and/or duration, but similar integrated amplitudes.
Thus, at late times
and high metallicities, all galaxies have the full complement
of N and thus the same N/O. (2) Star formation depends on the
size (e.g., mass, surface or volume density as in a Schmidt law) 
of the DLAs, and/or galactic winds.  
Hence, coeval but different systems will have 
had dissimilar star formation
histories and thus different N/O, with dispersion again decreasing
as the total star formation slows at late times.

The lesson we have learned in this section,
therefore, is that there is (very likely) intrinsic scatter
in the N--O data.  In contrast,
the small scatter in the helium abundances
is entirely consistent with the observational
errors.   Therefore, while
self-enrichment 
may well occur in BCGs, and might be responsible for the
N--O dispersion, the helium abundances cannot
(yet) confirm this scheme.  In view of this,
we will not attempt to model the observational
scatter in this paper; instead, we will limit ourselves
to models describing the mean abundance trends.
With this more modest goal, we concentrate on
the uncertainties inherent in chemical evolution
models of BCGs, and their implications for
deriving primordial helium abundances.

\section{The Model}
\label{sect:model}

To construct models for
BCG chemical evolution, 
we adopt the usual chemical evolution formalism,
allowing for outflows of both bulk ISM material
as well as enriched supernova ejecta.
For simplicity and in the absence of strong evidence otherwise,
we will adopt an initial mass function (hereafter, IMF) 
$\phi$, which is time-independent.  
Our normalization convention is 
$\int dm \, m \, \phi(m) = 1$
over the mass range $m \in (\minf,\msup)$.
Unless otherwise noted,
we will use 
the power-law (Salpeter) form, $\phi(m) \propto m^{-x}$.
The evolution of total baryonic mass $M_{\rm tot}$, gas mass $M_{\rm g}$, and 
mass fraction $X^i$ in species $i$, is given by
\beqar
\frac{d M_{\rm tot}}{dt} & = & - \outf \\
\frac{d M_{\rm g}}{dt} & = &  - \psi + E - \outf \\
\label{eq:x_i}
M_{\rm g} \, \frac{d X^{i}}{dt} & = &  E^i - X^i E 
    - ( {\vartheta}^{i} - X^i \outf)
\eeqar
where rate of outflow is $\outf$, and the rate of outflow in element $i$
is
${\vartheta}^{i}$; more on this below.
The rate of mass ejection from dying stars is, as usual,
\beq
E(t) = \int_{m(t)}^{\msup} \ dm \ m_{\rm ej}(m) \ \phi(m) \ \psi(t-\tau_m)
\eeq
where a star of initial mass $m$
ejects mass $m_{\rm ej} = m - m_{\rm rem}$, and
has a lifetime $\tau_m$ with inverse $m(t)$.
The mass ejected in species $i$ is
\beq
E^i(t) = \int_{m(t)}^{\msup} \  dm \ 
   m_{\rm ej}^{i}(m) \ \phi(m) \ \psi(t-\tau_m)
\eeq
where the nucleosynthesis yield
$m_{\rm ej}^{i}$ is the mass ejected in $i$, including
both unprocessed and new components.  Note
that the yields can depend on the metallicity.

We adopt the 
stellar evolution and nucleosynthesis
inputs as follows.
We use the $m_{\rm ej}$ of Iben \& Tutukov \pcite{it}, and
the stellar lifetimes $\tau_m$ of Scalo \pcite{scalo}.
As we will see, our results will be very sensitive to the choice of
stellar yields. Our ``fiducial'' set of yields
combines two recent studies which model in detail
the effects of advanced stellar evolution on
the final nucleosynthesis products.
For the nucleosynthesis of intermediate mass,
AGB stars ($1-8 \msol$)
we take the metallicity-dependent yields of
van den Hoek and Groenewegen \pcite{vdHG}.
The massive star ($11-40 \msol$) yields come from the metallicity-dependent
tabulation of Woosley and Weaver \pcite{ww}.
When stars of higher mass
(i.e., $> 40 \msol$) are included, we 
extrapolate these yields, again in a metallicity-dependent fashion.
Other important issues are whether to include
massive stellar winds (e.g., Maeder \cite{mae}),
and how to treat the uncertain 
``zone of avoidance'' of
progenitor masses in $8-11 \msol$.
Our fiducial case will not include the
effects of stellar winds, and will
interpolate the tabulated yields between $8-11 \msol$.
We will consider (\S \ref{sect:alt_yield}) 
alternatives to our fiducial case,
both by including stellar winds and different
$8-11 \msol$ schemes, and by adopting 
the wholly different yield tabulation of
Maeder \pcite{mae}.

The AGB yields of carbon and nitrogen depend
sensitively on the details of
hot bottom burning.  That is, nuclear burning
at the bottom of the convective envelope
can process C (and O) to N.  Thus, the
C and N yields depend sensitively on how
convection is modeled, e.g., in the
mixing length parameter $\alpha$.
The van den Hoek \& Groenewegen \pcite{vdHG}
models quantify hot bottom burning in terms
of $m_{\rm HBB}$, the core mass coordinate
at which the burning occurs.
They tabulate  results for their recommended value of 
$m_{\rm HBB} = 0.8 \msol$; they also give
results for 
$m_{\rm HBB} = 0.9 \msol$, which they
argue is an upper bound.
In view of these uncertainties, we will
treat $m_{\rm HBB}$ as a free parameter.
In a given model, we compute low-mass
yields by linear interpolation
between the two $m_{\rm HBB}$ tabulations
at a given metallicity:
\beq
m_{\rm ej}^{i}(m,Z) = 
  (1-\xhbb) \, m_{\rm ej}^{i}(m,Z;0.8) +
  \xhbb \, m_{\rm ej}^{i}(m,Z;0.9)
\eeq
where $m_{\rm ej}^{i}(m,Z;0.8)$ is the yield
for progenitor mass $m$, metallicity $Z$, and
$m_{\rm HBB} = 0.8$.
The dimensionless quantity $\xhbb$
parameterizes the interpolation between the tabulated values of 
$m_{\rm HBB}$.

In choosing a primordial helium abundance for our models,
we must address the $1\sigma$ discrepancy between the
$\yp$ intercepts derived from linear fits to N/H and to O/H
(Table \ref{tab:Y_fit}).
Since the nitrogen nucleosynthesis and hence
chemical evolution is less certain (in particular, the degree to which
there is secondary production of nitrogen is uncertain), we will adopt the
intercept, $\yp = 0.238$, from the
$Y-$O B98 fit.  
In comparing the models with the data, we will for the
most part focus on the slopes of the $Y-$N,O curves,
independent of the value of $\yp$.

With these inputs fixed from stellar evolution models, 
the central issue in building a model of BCGs is
the choice of an appropriate star formation rate
and outflow prescription.  Regarding the star formation rate,
observations indicate that BCGs are undergoing
bursts of star formation, and many of these systems have 
a history of stochastic bursts separated by
periods of quiescence.  In a simple 
closed box, instantaneous recycling
chemical evolution model, the details of the star formation
history are irrelevant for the abundance patterns.
However, the difference between bursting versus smooth
star formation can be important in more realistic models where:
(1) the assumption of instantaneous recycling is relaxed,
and finite stellar lifetimes are included, and 
(2) the assumption of a closed box is relaxed, and
outflows are adopted.  
Consequently, we will examine the effect 
of bursting versus smooth star formation rates; 
in fact, we will find that the effects
of bursting are not as important as one might at first
imagine.  

The collective effect of supernovae leads to
gas bulk motions, as well as heating and evaporation,
which we model the via outflows.
The outflow thus has two components.
(1) We allow a bulk ``ISM wind,'' with rate
$\outf_{\rm ISM}$, and the composition of the ISM:
$X_{\outf,{\rm ISM}}^{i} = X^i$.
For simplicity, we follow the Hartwick \pcite{hart}
model, in which the galactic wind is proportional
to the supernova rate (and hence star formation rate) that drives it:
\beq
\label{eq:wind_ISM}
\outf_{\rm ISM} = \effISM \psi
\eeq
Given that the outflow rates used here are relatively slow, this is
consistent with the prescription of Scully \etal \pcite{sean} based on
Larson \pcite{lar}.
As suggested by recent results of  Mac Low \& Ferrara \pcite{mlf}, the
mass ejection efficiency is likely to be low in BCGs if they have a
large complement of dark matter.
Note that the democracy of the ISM wind leads to
the absence of an explicit term in the abundance evolution.
However, these outflows can still be
influential as they reduce
$M_{\rm g}$ faster than in a closed box model.
(2) The other outflow component is an ``enriched wind,'' with rate 
$\outf_{\rm SN}$.  
This process posits the partial removal
of supernova ejecta and hence
heavy elements:
$X_{\outf,{\rm SN}}^{i} = X_{\rm ej,SN}^i = (E^i/E)_{\rm SN}$,
where in the last expression
the ejection integrals include only supernova ejecta.
We further assume that the ejection rate is proportional to
the total supernova mass ejection rate, with efficiency $\effSN$.
Thus we write 
$\outf_{\rm SN} = \effSN E_{\rm SN}$.  
In sum, therefore,
$\outf = \outf_{\rm ISM} + \outf_{\rm SN}$,
and $\outf^i = X^i \outf_{\rm ISM} + \effSN E^i_{\rm SN}$.
This gives, in eq.\ \pref{eq:x_i}
\beq
\label{eq:wind_SN}
({\vartheta}^i - X^i \outf) = 
   \effSN \left( E^i_{\rm SN} - X^i E_{\rm SN} \right)
\eeq
where the ISM term drops out because
$X_{\vartheta, {\rm ISM}}^i = X^i$.
As this wind is composed of only supernova products, it reduces the
effective supernova yield, and clearly will have a significant impact
on elements wholly or partially made in supernovae.

\section{Results}
\label{sect:results}

As the preceding section made clear, the
uncertainties in the parameter
and input physics choices leads to a large number
of possible models.  
We have investigated the parameter space
extensively.  
However, rather than presenting all models,
we will instead summarize the basic trends, and
show only selected results.

In order to judge a model's fit to the data,
we must consider how our one-zone, instantaneous mixing
model fits into a larger picture which acknowledges the
the presence of real scatter in N--O.
As pointed out by Pilyugin \pcite{pil}, 
well-mixed regions, have relaxed from
self-enrichment effects and lie on upper envelope of the N--O
data.  That is, 
points on the upper envelope of the N--O data
should come from systems that have had sufficient time
for low-mass star products to ``catch up'' with those
of high-mass stars (see discussion in \S \ref{sect:scatter}).  
Thus a good fit from our models should
lie close to this envelope (within errors), and not 
through the full data set.  On the other hand, since current data does
not support evidence for scatter in helium,  good models should fit the
central trends of helium versus N/H and O/H.

\subsection{Outflows and BCG Evolution}
\label{sect:outflow}

The goal of our study is to determine
which models can best fit
all of the HeNO data simultaneously and as we shall see, to identify
weak spots in the theoretical input which may be responsible for poorly
fit data.   In this section, we consider the case in which there are
outflows, with a standard set of nucleosynthesis yields (i.e., the
tabulations described in the \S \ref{sect:model}, with
linear interpolation
between AGB stars and supernovae in the
$8-11 \msol$ range).

A model thus requires specification of a star formation
rate, as well as four parameters:
(1) the IMF; (2) the strength of the bulk ISM outflow $\effISM$,
(3) the supernova ejection efficiency $\effSN$; 
and (4) the degree of hot bottom burning $\xhbb$.
Our strategy is as follows.
We will consider both smooth and bursting star formation
rates; in fact, we will find that the basic model
trends are not sensitive to the details of the
star formation rates.  For a given star formation
rate, we
try several different IMFs.   
For each IMF, we 
pick a range of $\effISM$ and $\effSN$.
Finally, we adjust the remaining parameter, $\xhbb$,
to get an acceptable fit to N--O (if this is possible).
The models that fit N--O we deem ``good candidates''
in that they fit the metal evolution, and we then
examine their helium evolution.

To summarize model results, we quote
average slopes $\Delta Y/\Delta A$.
We compute these slopes in a manner exactly analogous
to the fits to the B98 data set (Tables \ref{tab:Y_fit}
and \ref{tab:NO_fit}), as follows.  The models produce a series
of abundance points, one at each timestep.
We isolate each of the model points
within the B98 metallicity range:
$4 \le 10^7 {\rm N/H} \le 100$, and
$15 \le 10^{6} {\rm O/H} \le 150$
(the lower limits correspond to the lowest
measured metallicities, and the upper limits
are from the definition of the B98 set).
With these model points, we perform a least squares
linear fit to compute the slope and intercept.

We begin with the simple and instructive 
case of models with a smooth star formation rate.
Of course, BCGs demonstrate bursts of star formation,
the transient effects of which we will miss.
However, we will see that a smooth star formation rate
provides a good ``ensemble average'' of this stochastic behavior.  
For the smooth rate, we adopt $\psi = \lambda M_{\rm g}$,
with $\lambda = 0.29 \, {\rm Gyr}^{-1}$;
our results are insensitive to the value of $\lambda$.
We will examine these models with and without
bulk mass loss, i.e., with $\effISM$ both zero and nonzero;  in fact,
we will
find that the results are insensitive to $\effISM$.

We turn first to models with an IMF given by
$\phi(m) \propto m^{-2.7}$.
We will use this over a mass range 
$(m_{\rm inf},m_{\rm sup}) = (0.1 \msol, 80 \msol)$.
$\effSN$ is allowed to vary from 0 to 0.9 and
results appear in Table \ref{tab:std_IMF}.
We find a strong correlation between 
the allowed values of the enriched outflow
efficiency $\effSN$ and the hot bottom burning mass $\xhbb$.
This is a general feature of all of our models,
and one can understand it as follows.
We are searching for models which fit N--O at 
each value of $\effSN$. 
As $\effSN$ increases, more O is lost from the
BCG, while most of the N remains.  Consequently,  
the effective yield of O drops, and
N/O increases.  In turn, as $\effSN$ increases,
the needed $\xhbb$
must change to compensate by decreasing N.  
Note that $\xhbb = 0$
corresponds to a deeper hot bottom burning core 
($m_{\rm HBB} = 0.8\msol$), and hence a
higher temperature and more efficient processing,
than $\xhbb = 1$.
Thus $\xhbb = 0$ implies more
hot bottom burning, and more N production.
For N to decrease as $\effSN$ increases,
the value of $\xhbb$ must increase as well.
Indeed, the N/O behavior at $\xhbb \sim 1$
(minimal hot bottom burning)
sets the limiting value for our choice of
$\effSN$.\footnote{In fact, for illustration we consider
slight extrapolations to $\xhbb > 1$, i.e., 
$m_{\rm HBB} > 0.9\msol$.  However, van den Hoek \& Groenewegen
\pcite{vdHG} argue that $m_{\rm HBB} = 0.9\msol$ is an upper limit. 
Also,  the high-$\xhbb$ models in question keep so few metals
that they cannot reach the higher observed BCG metallicities
before processing most of their gas.}  We stress that the difference in
results in terms of N/O between $\xhbb = 0$ and $\xhbb = 1$ is quite
significant and reflects our uncertainty in the stellar yields in this
important stellar mass range.
Figure \ref{fig:hbb_dep} illustrates the effect of different
$\xhbb$ on the predicted N--O evolution.

\begin{table}[htb]
\caption{Models with IMF $\phi(m) \propto m^{-2.7}$}
\label{tab:std_IMF}
\begin{tabular}{|ccc|ccc|}
\hline\hline
\multicolumn{6}{|l|}{Smooth star formation rate:  
   $\psi = \lambda M_{\rm g}$}  \\
$\effISM$ & $\effSN$ & $\xhbb$ & $\Delta Y/\Delta Z$ & 
   $\Delta Y/\Delta {\rm O}$ & $\Delta Y/\Delta {\rm N}$  \\
\hline
0 & 0          & 0.6  & 0.95 & 21 & 340  \\
0 & 0.5        & 0.9  & 1.4  & 31 & 516  \\
0 & 0.75       & 1.05 & 2.8  & 62 & 975  \\
0 & 0.9$^{a}$  & 1.10 & 6.6  & 148 & 1281  \\
0.5 & 0          & 0.6  & 0.96 & 22 & 345  \\
0.5 & 0.5        & 0.9  & 1.4  & 32 & 555  \\
0.5 & 0.75$^{a}$ & 1.05 & 3.5  & 78 & 1111  \\
\hline\hline
\multicolumn{6}{|l|}{Bursting star formation:  
   $M_{\rm burst} = 0.15 M_{\rm tot}^0$} \\
$\effISM$ & $\effSN$ & $\xhbb$ & $\Delta Y/\Delta Z$ & 
   $\Delta Y/\Delta {\rm O}$ & $\Delta Y/\Delta {\rm N}$  \\
\hline
0 & 0          & 0.6  & 0.87 & 19 & 319  \\
0 & 0.5        & 0.9  & 1.1  & 25 & 447  \\
0 & 0.75$^{a}$ & 1.05 & 1.6  & 36 & 735  \\
0.5 & 0          & 0.6  & 0.89 & 20 & 354  \\
0.5 & 0.5$^{a}$  & 0.9  & 1.1  & 25 & 421  \\
0.5 & 0.75$^{a}$ & 1.05 & 1.6  & 35 & 694  \\
\hline\hline
\end{tabular} \\
${}^{a}$Gas content exhausted before ${\rm O/H} = 150 \times 10^6$.
\end{table}

\begin{table}[htb]
\caption{Models with Alternative Initial Mass Functions}
\label{tab:alt_IMF}
\begin{tabular}{|cc|ccc|}
\hline\hline
\multicolumn{5}{|c|}{$\phi \propto m^{-2.35}$}  \\
$\effSN$ & $\xhbb$ & $\Delta Y/\Delta Z$ & 
   $\Delta Y/\Delta {\rm O}$ & $\Delta Y/\Delta {\rm N}$  \\
\hline
0    & 0    & 0.70 & 16 & 233  \\
0.5  & 0.6  & 0.80 & 18 & 261  \\
0.75 & 0.95 & 1.1  & 24 & 449  \\
0.9  & 1.1  & 2.6  & 59 & 989  \\
\hline\hline
\multicolumn{5}{|c|}{$\phi \propto m^{-3}$}  \\
$\effSN$ & $\xhbb$ & $\Delta Y/\Delta Z$ & 
   $\Delta Y/\Delta {\rm O}$ & $\Delta Y/\Delta {\rm N}$  \\
\hline
0           & 0.85 & 1.9 & 43 & 679  \\
0.5${}^{a}$ & 1.0  & 3.6 & 79 & 1100  \\
\hline\hline
\multicolumn{5}{|c|}{$\phi \propto  m^{-1}
\exp[-\ln^2(m/m_c)/2\sigma^2]$}  \\
$\effSN$ & $\xhbb$ & $\Delta Y/\Delta Z$ & 
   $\Delta Y/\Delta {\rm O}$ & $\Delta Y/\Delta {\rm N}$  \\
\hline
0   & 0.7  & 1.09 & 24 & 333  \\
0.5 & 0.95 & 1.44 & 32 & 483  \\
0.7 & 1.07 & 2.22 & 50 & 856  \\
\hline\hline
\end{tabular} \\
${}^{a}$Gas content exhausted before ${\rm O/H} = 150 \times 10^6$.
\end{table}

The dependence on $\effSN$ on the slopes of Table \ref{tab:std_IMF} 
can also be understood physically in terms of the HeNO yields.
The simpler behavior is the increase of 
$\Delta Y/\Delta O$ with $\effSN$.
As the enriched outflows become stronger,
the supernova products are lost while
the AGB star products remain.
Oxygen is made exclusively by massive stars,
but helium is made roughly equally in both mass ranges,
so for all models which reach the same O/H in the BCG,
there is more new helium in those models with enriched outflows.
Thus $\Delta Y/\Delta$O increases with $\effSN$.
For nitrogen, the story is more complicated.
Since N comes mostly from intermediate-mass stars,
one might expect a behavior opposite to that of the
He slope versus O---i.e., $\Delta Y/\Delta {\rm N}$ 
should decrease.
However, this effect is outweighed by the 
increase in $\xhbb$ which is needed to offset the
changes in the N--O behavior due to the increasing $\effSN$.
As we raise $\xhbb$, the N yields are reduced, and
so $\Delta Y/\Delta {\rm N}$ increases.
Also, as noted by 
Pilyugin \pcite{pil}, 
the N yields are lowest at low metallicities;
so when the overall metallicity is suppressed at higher $\effSN \ga 0.75$,
$\Delta Y/\Delta {\rm N}$ increased.

Given the uncertainties in the IMF (particularly for
these extragalactic systems),   
we now examine the effect of a different IMF slope.
Table \ref{tab:alt_IMF} presents the results for 
the classic $\phi(m) \propto m^{-2.35}$, and
for the extreme case of $\phi(m) \propto m^{-3}$.
We also present results for
a log-normal form 
$\phi(m) \propto m^{-1} \exp [-\ln^2(m/m_c)/2\sigma^2]$, 
which has $m_c = 0.1 \msol$ and $\sigma = 1.6$,
and is similar to that of
Miller \& Scalo \pcite{ms}, Scalo \pcite{scalo}, and
Kroupa, Tout, \& Gilmore \pcite{kroupa}.
Figure \ref{fig:imf_dep} illustrates these results for
$\effSN = 0.5$.
In all cases, we take $(m_{\rm inf},m_{\rm sup}) = (0.1 \msol, 80 \msol)$.
A comparison with the
previous results shows that,
all other parameters being equal, the steeper IMF slope gives
larger helium slopes.  This effect is closely
related to the increase of the slopes with $\effSN$,
since steepening the IMF is equivalent to removing
supernova ejecta.  Note that other than larger overall
slopes, the basic trends are the same as for 
the less steep IMF.

For all adopted IMFs,
we find that the helium slopes do increase with the 
enriched outflow
strength, but never become as large as the
slopes in the data fits.
This trend is common to most models we have explored.
Indeed, we can reproduce the observed
slopes only a few cases with alternative yields,
discussed below (\S \ref{sect:alt_yield}).
However, we can already make some conclusions
regarding the behavior of models with
the fiducial yields we have adopted.  
First, these models predict a N--O evolution that
is close to linear, and $Y$ evolution that
is also very close to linear 
in both O and N.  It is thus clear that
chemical evolution models can support
that linear behavior that is the simplest interpretation
of the data, and has been used in the fits
to derive \yp.  Since these models do not reproduce the
slopes quantitatively, one cannot infer that a
successful model would {\em demand} linearity,
but it is at least clear that linear evolution of
$Y$ versus metals is the typical result of a broad set of
models.  

Since our models give very linear $Y-$N and $Y-$O trends,
it also follows that these
models do not predict a significant difference
in slope between the lower and higher metallicity
BCG points.  That is, these models do not predict
a difference between the slopes in the B versus C data sets.\footnote{This 
is not {\em per se} a liability of the models,
as the data at present only weakly suggest that the
B and C slopes might differ, as one can see from Table 1, when errors in
the slopes are taken into account.} The linearity of models' $Y-$N,O
further implies that the models do not predict that the extrapolated
intercepts---$\yp$(N) and $\yp$(O)---should differ
significantly from the true $\yp^{\rm true}$.  Indeed, we find
this to be the case, not just for oxygen but also for nitrogen:
the errors one expects from linear extrapolations are small,
$\yp({\rm N})-\yp^{\rm true} < 0.001$ for all models.
This is in contrast to the data, for which
$\yp({\rm N}) - \yp({\rm O}) = 0.002-0.003$.
Our results thus differ from those of Fields \pcite{fields},
who found that $\yp({\rm N})-\yp^{\rm true}$ as large as 
$0.005$ for models in which an explicitly secondary 
source of N was put in by hand.
Here, however, we have adhered to the tabulated
van den Hoek \& Groenewegen \pcite{vdHG} yields, which
include hot bottom burning, and thus some some degree
of secondary N.  We find, however, 
that the secondary N component in our results is weak.  
Furthermore, the small $\Delta Y/\Delta$N slopes
in our models force the extrapolated $\yp$(N) to
be accurate, independent of any effects due to secondary N.

Although 
the model slopes do fall below 
the data regressions, 
it is nevertheless noteworthy that
the model curves 
run through most
of the data's error bars---i.e., the $\chi^2$ is not bad.
Thus, while these models are not good fits to the
data, they are not strongly excluded by the data either.
This comes about because the observational errors in the individual
helium abundances are large compared to the
change $\Delta Y$ over a small baseline in metallicity.
As a result, even a small slope cannot be strongly
ruled out unless one considers higher metallicity points,
where there is more leverage to compare the slopes.
Thus, the BCG observational ``wish list'' 
for addressing these chemical evolution issues is some different than 
that for obtaining primordial helium.  Chemical evolution
is best constrained by data over a large range of metallicities,
whereas primordial helium is best obtained via the systems
with the lowest possible metallicities.

We should note at this point that the He data of 
Izotov \& Thuan \pcite{it1}, based on the observation
of 5 He lines, does indicate a smaller $\Delta Y/\Delta$O slope,
as mentioned in \S \ref{sect:data}.  As one can see from the tables,
a slope of $\Delta Y/\Delta {\rm O} = 44 \pm 19$ is not hard to
accommodate in any of the models considered.
Although we believe the uncertainties with this approach
to be large, we should not exclude the possibility that
the discrepancy in the slopes has its source in the He data.

With these caveats in mind, we identify the best models
having our fiducial yields.
The best model with 
the $m^{-2.7}$ IMF has $\effSN = 0.75$ (if we
include carbon evolution as a constraint, 
$\effSN = 0.5$ is the maximum allowable).
The slopes in this model are $\Delta Y/\Delta O = 62$,
and $\Delta Y/\Delta N = 975$ which respectively
are $2\sigma$ and 
$2.1 \sigma$ away from the data regressions.
The favored value of $\effSN \simeq 0.75$ comes about
as follows.
A larger $\effSN$ would lead to larger helium slopes
more in line with the data.  However, as $\effSN \rightarrow 1$,
the systems loses so much of its heavy element products that 
it no longer can achieve metallicities observed in BCGs
(e.g., for $\effSN=0.9$, when the gas fraction dips below 5\%, 
O/H is only $80 \times 10^{-6}$;
observed values are as high
as ${\rm O/H} \sim 150 \times 10^{-6}$). We note that
$\effSN = 0.75$ is similar to values  obtained in other studies of BCG
models, and in some models of Galactic outflows.
Also, this value is in line with those 
found by Mac Low \& Ferrara \pcite{mlf}.
The only models significantly better than this
are the ones with the steepest IMF, $\propto m^{-3}$, 
We note that these models require that the
enriched outflow efficiency must be very small
$< 0.5$ in order to reproduce the observed BCG metallicities.
Further, these models have trouble
reproducing the highest observed metallicities,
and have C abundances which 
are much larger than the available observations (see below, this section).

It is of interest to point out model parameters that
do {\em not} have a strong impact on the results.
One is the character of the 
star formation rate.  As seen in Table \ref{tab:std_IMF} 
the results are quite insensitive
to whether a smooth or bursting star formation rate is used.
This is not to say that the bursting star formation rates
give the same results in detail.  As seen in 
Figure \ref{fig:sfr_dep}, the bursting star formation
leads to a ``stairstep'' evolution in the N--O plane
in contrast to the smooth curve for the smooth star formation
rate.  While the curves are locally very different,
they are globally very similar---the bursting trend
is a variation around the smooth one.  In particular,
the smooth evolution lies on the upper envelope of the
bursting one.  This justifies the self-consistency of
our approach to fitting the data:  we have assumed that the N--O
scatter is real and due to self-enrichment effects of bursting, 
so we have fit the upper envelope of the data.
Furthermore, since different BCGs undergo different numbers
and magnitudes of bursts, a smooth star formation rate 
can be seen as an ``ensemble average'' over
different bursting histories.
This simple, smooth star formation rate
is valid long as one is careful to
fit the upper envelope of the N--O data.  
We note that these issues are not relevant for the He evolution;
there, the differences between the bursting and smooth evolution
curves are negligible.
Our conclusion in these regards thus
agrees with those of Pilyugin \pcite{pil,pil96}.

Even less important for our results is the
bulk ISM outflow.  As seen in Table \ref{tab:std_IMF},
the results for $\effISM = 0$ and $\effISM = 0.5$
are essentially indistinguishable for all but a few cases.
This is due to the democratic character of the bulk
outflow, whose composition is the same as that of
the ISM.  Indeed, in the 
instantaneous recycling approximation,
such an outflow leads to a uniform reduction in
the effective yields by a factor $1+\effISM$,
and thus there is no effect in element {\em ratios} or
the abundance--abundance trends of interest here.
We see here that even without instantaneous recycling,
$\effISM \ne 0$ only has an impact when
$\effSN$ is large.  In this case, the extremely reduced
effective yields feed back to the metallicity dependent yields,
and so modify the helium slopes.

Finally, we turn to the evolution
of carbon in our models.
Carigi et al.\ \pcite{ccps} noted
the importance of C/O as a 
constraint on the high-mass stellar
cutoff $m_{\rm sup}$, and on
enriched winds.  Here we note the importance of
C in constraining intermediate-mass nucleosynthesis.
Figure \ref{fig:carbon} plots
C versus O, and $Y$ versus C,
for our best HeNO models at each IMF slope.
Note that C is high in models with high
$\xhbb$, i.e., when hot bottom burning is reduced.  
This comes about due to a tradeoff between
C and N---hot bottom burning produced N
at the expense of C.  Thus, the models that
have less hot bottom burning (to improve the fit to the N--O data)
also have high carbon, often above the C--O data.
A good compromise appears to be around
$\xhbb = 0.5$ for both power-law IMFs.
Thus, the C--O data prevents one from adopting
arbitrarily little hot bottom burning (i.e., large $\xhbb$);
furthermore, $\xhbb$ correlates with $\effSN$
in acceptable HeNO models, and so
the C--O effectively constrains the level of enriched
outflows.
Carbon can be a very strong and useful 
constraint on BCG models.

The carbon data generally support the idea of self-enrichment as 
a source of the N--O scatter.
Kobulnicky \& Skillman \pcite{ks},
measure C/H in three BCGs which have roughly
the same O/H but 
different N/O.  
Kobulnicky \& Skillman \pcite{ks}
find C/H variations, which are correlated with N/O--i.e.,
these galaxies have similar N/C ratios. 
This suggests that the nucleosynthesis of these
two elements is linked--as
we indeed expect:  both come from intermediate-mass stars.
Furthermore, this result is
consistent with the self-enrichment 
(``delayed release'' in the parlance of Kobulnicky \& Skillman)
scheme for the N--O scatter (\S \ref{sect:scatter}).
Complicating this picture are the results of
Garnett et al.\ \pcite{gsds}.  They measured 
C/H in the very low metallicity 
I Zw 18, and found 
$\log ({\rm C/O})_{\rm I \, Zw \, 18} \simeq -0.6 \pm 0.1$
which gives [C/O]${}_{\rm I \, Zw \, 18} = -0.2$.
This is high compared to other low-metallicity BCGs
($\log {\rm C/O} \simeq -0.85$), and much
higher than massive star models predict.
Garnett et al.\ interpret this as evidence for prior star formation 
and ejection of intermediate mass star products.
If so, then in the self-enrichment 
picture the high C/O ratio suggests that the
previous star formation has now returned all of
its ejecta, and so this C/O is more representative
of the smooth star forming average.  

We conclude this section by noting that
the trends in
our results (e.g., the effect of 
bulk and enriched outflows, or of the IMF)
are in broad agreement with 
the recent work of Pilyugin \pcite{pil}; Matteucci \& Tosi \pcite{mt};
Carigi et al.\ \pcite{ccps}.  
It is of particular interest to compare our results to
the recent models of 
Pilyugin \pcite{pil}.  
His models employ differential stellar winds, 
and are more detailed than ours in that 
they distinguish the abundances in the bulk
BCG gas and the star forming \hii\ regions.
Pilyugin thus models the effects of self-enrichment, 
and finds that while the predicted N--O scatter is large,
the $Y-$N,O scatter is smaller, in line with our results.

\subsection{The Effect of Alternative Yields}
\label{sect:alt_yield}

It was noted in (Fields \cite{fields}) that 
although $8-11 \msol$ yields 
are uncertain, they could have a large effect on helium
production.  In particular, it is unclear what fate
meets stars born in these objects.  Stellar evolution
models for this mass range are 
uncertain and are difficult to construct.
Indeed, modern tabulated yields
for either AGB stars or for type II supernovae
generally omit this mass range.
While it is considered likely that $8-11 \msol$ objects explode
as electron-capture supernovae (Nomoto \cite{nomoto};
Iben, Ritossa, \& Garc\'{\i}a-Berro \cite{irgb}),
the composition of the ejecta is unclear 
(e.g., Wheeler, Cowan, \& Hillebrandt \cite{wch}
propose these objects as the site of the r-process).
Woosley \& Weaver \pcite{ww86} suggested that
these objects might be supernovae
whose predominant product is helium, with very small metal yields. 
Indeed, this is
the trend down to the limits of the core collapse models.
If this is the case, it could significantly increase the
helium yield, given the large IMF weighting of this mass range
compared to that of the better known, high mass supernovae.  

To investigate this possibility,
we adopt a modified form of the approximation of 
Fields \pcite{fields}.  Namely, we assume
8 \msol\ is the lower limit for supernova production
(this is important in determining the wind strength
via eqs.\ \ref{eq:wind_ISM} and \ref{eq:wind_SN}).
We assume the yield at 8 \msol\ is ``all'' \he4,
i.e., any portion of the helium core that is ejected
contains only \he4, while the envelope remains that
of the ISM at the star's birth.  We fix the
yield at 11 \msol\ as that of Woosley \& Weaver \pcite{ww}.
Between these extrema, we determine
the yields via linear interpolation.
This differs from the procedure of section \S \ref{sect:outflow}, 
which uses the 8 \msol\
van den Hoek \& Groenewegen \pcite{vdHG} yields in the interpolation.
Thus our modification trades a decrease in C and N for an increase in He.
We see by comparing the results of Table 5 with those of Tables 3 and 4, 
that adding enhanced $8-11
\msol$ He production does increase the slopes, but only mildly.
Though the yield modifications which give rise to
this portion of
Table \ref{tab:alt_yields} are, in our opinion, plausible, and
help ameliorate the disagreement between observation and theory.
These changes are not, however, the solution to the problem.

\begin{table}[htb]
\caption{Models with alternative yields.}
\label{tab:alt_yields}
\begin{tabular}{|ccc|ccc|}
\hline\hline
\multicolumn{6}{|c|}{Modified yields from $8-11\msol$ stars} \\
$\effISM$ & $\effSN$ & $\xhbb$ & $\Delta Y/\Delta Z$ & 
   $\Delta Y/\Delta {\rm O}$  & $\Delta Y/\Delta {\rm N}$ \\
\hline
0 & 0   & 0.5  & 1.0 & 23 & 343  \\
0 & 0.5 & 0.85 & 1.5 & 34 & 526  \\
0 & 0.75 & 1.0  & 3.3 & 73 & 896  \\
\hline\hline
\multicolumn{6}{|c|}{Maeder \pcite{mae} yields} \\
IMF & $\effSN$ & $\xhbb$ & $\Delta Y/\Delta Z$ & 
   $\Delta Y/\Delta {\rm O}$ & $\Delta Y/\Delta {\rm N}$  \\
\hline
$\phi \propto m^{-2.35}$ & 0    & N/A & 0.84 & 19  & N/A  \\
$\phi \propto m^{-2.7}$  & 0    & N/A & 1.5  & 24  & N/A  \\
log-normal               & 0    & N/A & 1.9  & 42  & N/A  \\
log-normal               & 0.5  & N/A & 3.3  & 74  & N/A  \\
log-normal               & 0.75 & N/A & 6.1  & 137 & N/A  \\
log-normal               & 0.9  & N/A & 14   & 320 & N/A  \\
\hline\hline
\multicolumn{6}{|c|}{$m_{\rm sup} = 40 \msol$; $\phi \propto m^{-2.7}$} \\
$\effISM$ & $\effSN$ & $\xhbb$ & $\Delta Y/\Delta Z$ & 
   $\Delta Y/\Delta {\rm O}$ & $\Delta Y/\Delta {\rm N}$  \\
\hline
0 & 0   & 0.8 & 1.4 & 31 & 483  \\
0 & 0.5 & 1.0 & 2.2 & 50 & 802  \\
\hline\hline
\end{tabular}
\end{table}

We have also considered the effect of
uncertainties in our adopted yields over the whole range of masses,
by considering an alternate yield tabulation. 
Traat \pcite{traat} pointed out that
Maeder's \pcite{mae} yields can give large He slopes,
but only when used in combination with
Scalo's \pcite{scalo} IMF.
To investigate this claim, we replaced
our fiducial yields with those of Maeder \pcite{mae};
these include stars from $0.8-120 \msol$.
These yields are based on the Schaller et al.\ \pcite{schal} models
go up to the early asymptotic giant branch
(E-AGB) phase for intermediate-mass stars, and
up through central C burning for massive stars.
Thus, these models do not provide a detailed treatment of
AGB evolution and nucleosynthesis in intermediate-mass stars, 
nor of the advanced and explosive evolution of massive stars.
In fact, the Schaller et al.\ \pcite{schal} models, that 
are the basis of
the Maeder \pcite{mae}, 
provide the initial conditions for the AGB studies of 
van den Hoek \& Groenewegen \pcite{vdHG}.    

Since Maeder \pcite{mae} does not provide N yields,
a complete comparison with our fiducial yields is impossible.
Nevertheless, we adopt the Maeder yields
for He and O, which are tabulated for $Z = 0.001$ and
$Z = 0.02$.
As seen in Table \ref{tab:alt_yields} and in Figure \ref{fig:yield_dep},
the slopes for power law IMFs are somewhat increased 
relative to our fiducial models, but the effect 
is relatively small---not 
enough to bring agreement with the data.
However, in the case of a log-normal IMF, the 
change is indeed large, as discussed by
Traat \pcite{traat}, and in contrast to
our fiducial case which does not strongly change
with this IMF.  We find that slopes now come within
$< 2\sigma$ of the data regressions, in the case
without enriched winds ($\effSN=0$).  
The slope increases when enriched winds are invoked,
and for $\effSN = 0.5-0.75$, 
the slopes lie comfortably in the range of observations.
While this increase is tantalizing, it is not
so clear that this puts and end to the slope problem.
It is troubling that models which treat the advanced phases of
stellar evolution in a more detailed way are unable to
reproduce the large slopes.  At any rate, this 
points up: (1) the sensitivity of our results to yields, 
(2) the need for accurate nucleosynthesis models, and
(2) the likelihood
that a problem with yields lies at the heart of the slope problem,
so long as the slopes inferred from the data are taken at face value.

We have also considered the effect of massive star winds
on our models.  
Following Fields \pcite{fields},
we used the Maeder \cite{mae} tabulation of
stellar wind yields (which do include N) in conjunction with our fiducial
massive star yields of Woosley \& Weaver \cite{ww}.
We found that in fact, the stellar winds slightly 
lowered the helium slopes, since increase in He production
more than offset by increase in N and O.
This conclusion is consistent with that of Fields \pcite{fields}.
This is perhaps not a surprise since the effects of winds are generally
small at the low values of $Z$ we are considering.

Another proposal to increase the predicted 
helium slopes appeals to a change in yields
via an effective change in the IMF mass limits.  
Namely, the oxygen yields are reduced if one
lowers $m_{\rm sup}$, the upper mass limit
for stars formed.\footnote{We have also run models
with $m_{\rm sup} = 80 \msol$ but an increased
$m_{\rm inf} = 0.3 \msol$.  This change has only
a modest effect, in the direction of further lowering
the He slopes.}
Following Olive, Thielemann, \& Truran \pcite{ott},
Maeder \pcite{mae} has suggested that
while stars may still be formed up to $m_{\rm sup} = 80$
or even higher, but that stars more massive
to some cutoff $m_{\rm BH}$ will become
black holes during core collapse, and will
not return their nucleosynthesis products.
For our purposes, adopting a $m_{\rm BH} < m_{\rm sup}$
is just equivalent to lowering $m_{\rm sup}$.
We have run models with $m_{\rm sup} = 40 \msol$;
results appear in Table \ref{tab:alt_yields}.
Because high mass star yields are reduced,
a good fit to N--O requires less hot bottom burning 
(i.e., higher $\xhbb$) than in models with our fiducial 
$m_{\rm sup} = 80 \msol$.
We find that $\xhbb = 1$ is required for $\effSN = 0.5$;
for the case, the change in the mass limit does indeed
raise the helium slopes, to 
$\Delta Y/\Delta {\rm O} = 47$, and $\Delta Y/\Delta {\rm N} = 805$. 
These are in marginal agreement with the data.
(Lowering $m_{\rm sup}$ further to $30 \msol$
does not help: the $\effSN=0$ slopes
are $\Delta Y/\Delta {\rm O} = 41$ and $\Delta Y/\Delta {\rm N} = 626$, 
but the N--O data already requires the minimum
hot bottom burning, $\xhbb = 1$, so enriched outflow
cannot be invoked to improve these results.)
However, as noted at the end of \S \ref{sect:results},
the low degree of hot bottom burning in this model
is unable to reproduce the C--O relation.
If the carbon data are to be taken seriously 
(and certainly if they improve) then 
one has to adopt a lower $\xhbb$, and hence a lower
$\effSN$, and this scenario become ineffective
in generating large helium slopes.
This underscores the power of carbon in particular,
and multiple element abundances in general, to constrain
BCG chemical evolution models.

Finally, we note that issues of stellar yields
probably underlie a notable contrast between our
findings and those of Pilyugin \pcite{pil}.
Pilyugin's calculations, unlike ours,
are able to achieve large $\Delta Y/\Delta$N
and $\Delta Y/\Delta$O while still fitting
N--O.  We strongly suspect that this crucial difference 
stems from  our different choices for massive stellar yields.
Pilyugin adopts the yields of
Chiosi \& Caimmi \pcite{cc},
which themselves are re-analyses of
older tabulations, corrected to
reflect the effects of a very strong mass loss rate.
Chiosi \& Caimmi \pcite{cc} argued
that strong mass loss should occur in essentially
all massive stars, and thus readjusted
the relation between the He core mass
and the progenitor mass to reflect this and to
rescale the yields.
The upshot of this rescaling is that 
stars of a given initial mass lose more material, and
so have smaller He cores and thus smaller heavy
element yields.
For example, Chiosi \& Caimmi \pcite{cc} estimate that 
the oxygen yield of a $23 \msol$ star 
is reduced from $2.0 \msol$ in a model without winds,
to $0.17 \msol$ with winds.  Clearly, a reduction this sharp
leads to a dramatic increase in $\Delta Y/\Delta$O
(or equivalently, $\Delta Y/\Delta Z$),
as Pilyugin indeed finds.
However, there is not evidence for such a 
strong effect in more modern supernova codes:
neither in Woosley \& Weaver \pcite{ww}, nor even
in the models of Maeder \pcite{mae}, who includes
stellar winds but finds
their effect to be  small below $\sim 25 \msol$.

\subsection{On Galactic and Extragalactic $\Delta Y/\Delta Z$}

The helium slope versus metallicity, $\Delta Y/\Delta Z$,
has long been viewed as a significant test of chemical evolution
models of BCGs as well as our own Galaxy.
It is indeed worrying that
models have consistently had trouble reproducing the observed
slopes in either of these systems since generic models tend to produce 
$\Delta Y/\Delta Z \sim 1$. 
Our preceding discussion has compared the 
BCG helium slopes as observed and as calculated our model;
here we discuss the relation of this work to
the situation in our Galaxy.

Studies of $\Delta Y/\Delta Z$ in our Galaxy
have focussed on 
the ``fine structure'' in the main sequence  of nearby stars
(e.g., Pagel \& Portinari \cite{pp}, and references therein).
This analysis notes the dependence of stellar luminosity,
at fixed effective temperature, on the composition.
Using HIPPARCHOS parallax data to obtain accurate luminosities,
Pagel \& Portinari \pcite{pp} infer
$\Delta Y/\Delta Z = 3 \pm 2$. 
For comparison,
solar system determinations of metallicity (Anders \& Grevesse \cite{ag})
and protosolar helium (Bahcall \& Pinsonneault \cite{bp}) would
imply a 
$\Delta Y/\Delta Z \sim 2$ (adopting $\yp = 0.238$).
These Galactic determinations are below
those from BCGs, though the uncertainties are large.  
If the slopes for these two systems can be shown to differ,
it could be a consequence of the different environments
and star forming histories of these systems.  
The connection one expects between the Galactic and extragalactic 
observed slopes depends on how
one models the chemical evolution of both systems.  
For example, if one favors BCG models which invoke enriched
outflows, but Galactic models that do not, one indeed expects a slope that is 
smaller for Galaxy.
If enriched outflows are not important for BCGs, or 
are present in our own Galaxy with similar efficiencies, than
the Galactic and extragalactic slopes should be the same
(although the Galactic data is for stars with higher metallicity
than in BCGs).

On the theoretical side, our results have implications for
Galactic chemical evolution
models.  We find that standard IMFs and 
up-to-date yields generically give $\Delta Y/\Delta Z \simeq 1$,
including those models without 
enriched winds ($\effSN=0$).  
Thus, even if the Galaxy had no such winds, this basic result
obtains.  Adding primordial infall does not affect the situation 
to first order, though the overall suppression of metallicity
would somewhat lower the N yields and thus augment
$\Delta Y/\Delta$N (without changing $\Delta Y/\Delta$O appreciably).   
Thus, it appears that the IMF and yield choices we have adopted
may have trouble with the 
the Galactic helium slopes ($\Delta Y/\Delta Z \sim 3$),
as well as those of BCGs.

We note in passing that 
the small scatter in the
$Y-$N,O trend could point to the uniformity of the
enriched wind efficiency 
over different BCGs.
That is, the
slope $\Delta Y/\Delta$O  is very sensitive to $\effSN$;
if different systems had very different $\effSN$, then
at a fixed O/H, one would expect a large range of 
$Y$, and thus the $Y - {\rm O}$ trend would have a large
intrinsic scatter.  The fact that it does not suggests
that $\effSN$ does not vary much.

\section{Discussion and Conclusions} 

We have considered the chemical evolution of
BCGs, through an analysis of the observed
HeNO abundances, and by studying chemical evolution
models.  Our basic conclusions are as follows.

The scatter in the $Y-$N and $Y-$O data is entirely consistent
with the observational errors.  In particular,
these data do not show the correlations (in the N--O plane)
that one would expect if the scatter were due to
self-enrichment by massive star ejecta.  On the other
hand, it appears that the N--O scatter real.
We expect that at least some of the observed scatter 
is due to self-enrichment, as is also suggested by
the observed correlation of high C/O systems with high N/O.

Our study of chemical evolution models of BCGs
concludes that  the favored models for these
systems (i.e., including outflows), coupled with
recent and detailed stellar yield calculations
(which are metallicity dependent), 
are not successful in reproducing simultaneously
all of the abundance trends.
We do find that in general, the models do
predict  
linear slopes in He versus N and O,
as has long been assumed in phenomenological fits to the data. 
The large slopes are, however, hard to reproduce 
quantitatively---even in models with supernova-enriched outflows.  
The only models which {\em can} produce large slopes,
while (possibly) maintaining agreement with 
N--O data, are those using the 
(less detailed) stellar yields
of Maeder \pcite{mae} in combination with
a log-normal IMF.  Consequently, we suspect
that the root of the problem lies in the nucleosynthesis inputs,
particularly the He yields.  We emphasize that
any means of improving the helium slopes
must be tested in models which fit {\em all} of HeNO,
and preferably C as well.

Another shortcoming of the models regards nitrogen.
We find that the models with the detailed yields
typically do not predict a strong enough secondary character 
for N versus O.  As a result, they do not find
that the $\Delta Y/\Delta $N slopes 
change much with metallicity.  In particular,
the models do not reproduce 
the differences found in the $Y-$N
slopes (and intercepts) between the full
and low-metallicity data sets.

Despite these unresolved issues, our study of BCG models 
does allow us to draw some conclusions.
Most models with detailed yields predicted
$Y-$N,O relations that were close to linear;
this gives some theoretical justification for
the adoption linear regressions of the data
(particularly for $Y$--O).  
Also, we confirm that the nature of the star formation rate---bursting 
versus smooth---does strongly affect the detailed evolutionary
history, and an ensemble of different histories
leads to significant scatter in N--O.  Fortunately, however, 
the N--O evolution 
in the smooth star forming history forms an
upper envelope to the N--O trends in the bursting models.
Thus, if one is not interested in the scatter, one
can adopt a smooth star forming model as long as
one is careful to fit appropriate envelope of the data.
We have also shown that carbon data 
provides an important additional constraint
on BCG chemical evolution.  In particular, C and N together
can diagnose the tradeoff between C and N
controlled by hot bottom burning. 
Additional C observations would be very useful in
further constraining BCG models.

Finally, aspects of our results have implications for
systems other than BGCs.
The difficulty for our models to reproduce the helium slopes
probably carries over to our own 
Galaxy.  Although there is some evidence that
the Galactic $\Delta Y/\Delta Z$ may be lower than in BCGs,
it nevertheless seems to exceed the values ($\sim 1$)
consistently predicted by all of our models
without strong enriched winds.
A good solar neighborhood $\Delta Y/\Delta Z$ would help
illuminate whether the helium slope problem is due to yields, or due
to environmental differences between the Galaxy and BCGs.
Finally, we note the similarity between the 
low metallicity N--O evolution in
BCGs, QSO absorbers (c.f. Lu, Sargent, \& Barlow \cite{lsb}),
and other external galaxies
(van Zee, Salzer, \& Haynes \cite{vzsh}).
The observed trends are compatible with each other, at least
to a first approximation.  This suggests that
the N--O evolution of these objects is similar, which 
may then imply that the enriched outflow of BCGs is
small.  If so, this is additional evidence that the
heart of the problem of the low helium slopes lies in the
stellar yields themselves.

\acknowledgments
We are grateful for discussions with
Sean Scully, Evan Skillman, Gary Steigman, and 
Don Garnett.  We particularly remember and acknowledge 
David Schramm, whose thinking on these issues
has profoundly influenced our own, and
whose friendship, enthusiasm, and insight we will deeply miss.
This work was 
supported in part by
DoE grant DE-FG02-94ER-40823.

\nobreak

\newpage

\begin{center}
{\bf FIGURE CAPTIONS}
\end{center}

\begin{enumerate}

\item \label{fig:NOcut}
{\bf (a)} Nitrogen versus oxygen for 62 BCGs;
data are from the OSS97 compilation (set B).
The solid line is the quadratic fit
with parameters from Table \ref{tab:NO_fit}, and divides the data: 
solid points have central values above the fit (``high N''),
open points below (``low N'').   \\
{\bf (b)}  The points of part (a), viewed in the
$Y-$O planes.  Note that the high and low
N points each fill the scatter about equally at fixed
metallicity.  See discussion in text.  \\
{\bf (c)} As in (b), for the $Y-$N plane. 

\item \label{fig:hbb_dep}
N versus O for different $\xhbb$.
IMF $\propto m^{-2.7}$, and $\effSN = 0.5$.

\item \label{fig:imf_dep}
The $Y-$N,O for different IMFs, with $\effSN = 0.5$.
Shown are curves for
$\phi \propto m^{-x}$, with
$x=2.7$ (solid line),  $x=2.35$ (dotted line),
$x=3$
(short dashes), and for
a log-normal IMF (long dashes).
Error bars in the data have been suppressed for clarity.

\item \label{fig:sfr_dep}
The effect of the star formation rate on
{\bf (a)} the N--O relation, and {\bf (b)} the $Y-$O relation.  Models
have $\effSN = 0$.  The solid curve is for
a smooth star formation rate, $\psi = \lambda M_{\rm g}$;
the broken curve is for a series of star formation bursts,
for which a gas mass $M_{\rm burst} = 0.15 M_{\rm tot}^0$ is processed
in each burst.  
Error bars in the data have been suppressed for clarity.

\item \label{fig:carbon}
The evolution of carbon in models with
fiducial yields.  The solid curve is for
the $m^{-2.7}$ IMF with $\effSN = 0.75$;
the broken curve is for 
the $m^{-2.35}$ IMF with $\effSN = 0.75$.
{\bf (a)} Carbon versus oxygen.  {\bf (b)} Helium versus carbon.

\item \label{fig:yield_dep}
The $Y-$O relation for different yields, with $\effSN$ as indicated.
For clarity, we show the $1-\sigma$ envelope of the
best-fit data regression, rather than the data points themselves.

\end{enumerate}


 \newpage

\begin{figure}[htb]
\vskip 1in
\hskip 1in
\epsfysize=7truein
\epsfbox{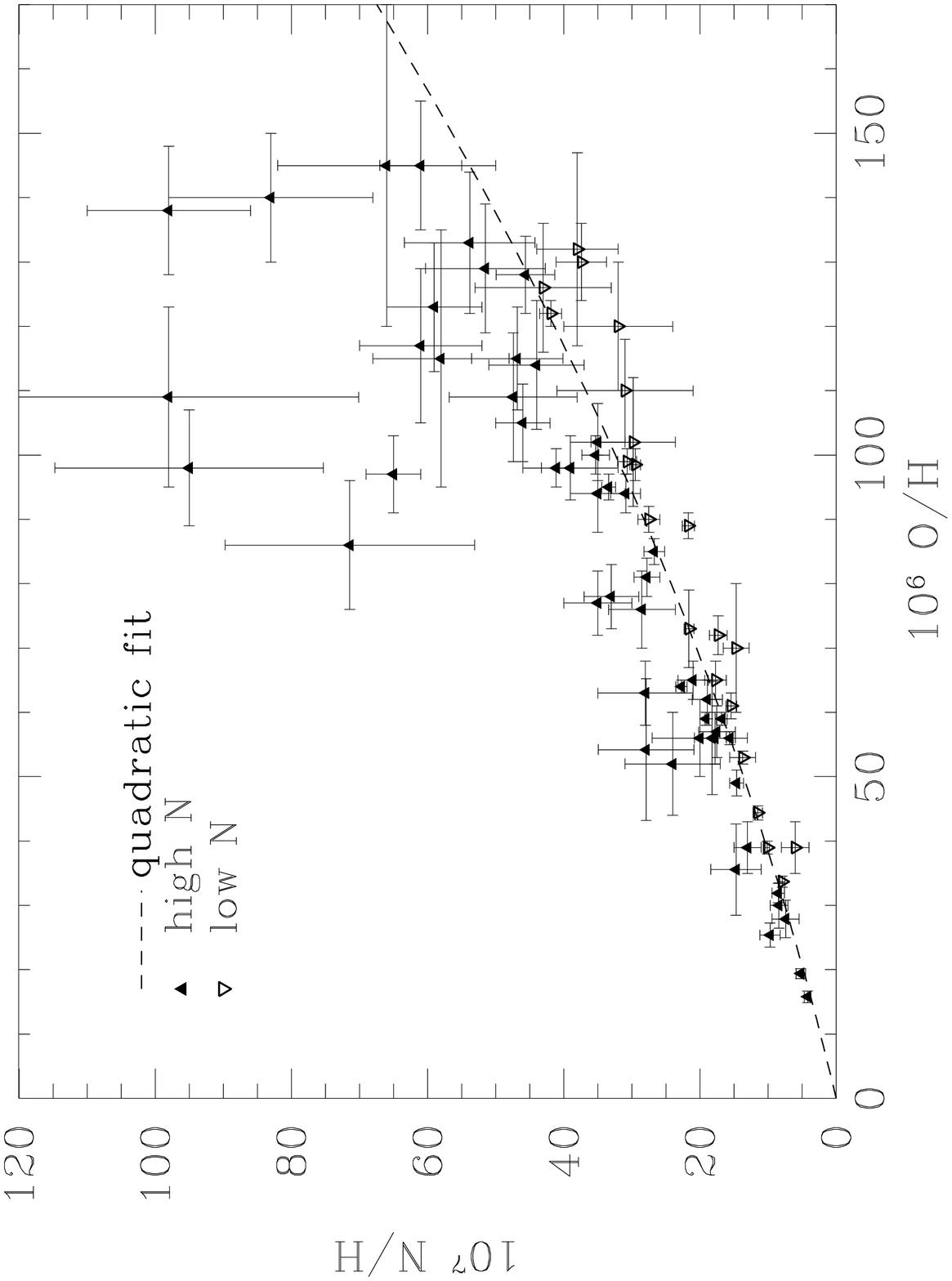}
\end{figure}  
\newpage

\begin{figure}[htb]
\epsfysize=7.5truein
\epsfbox{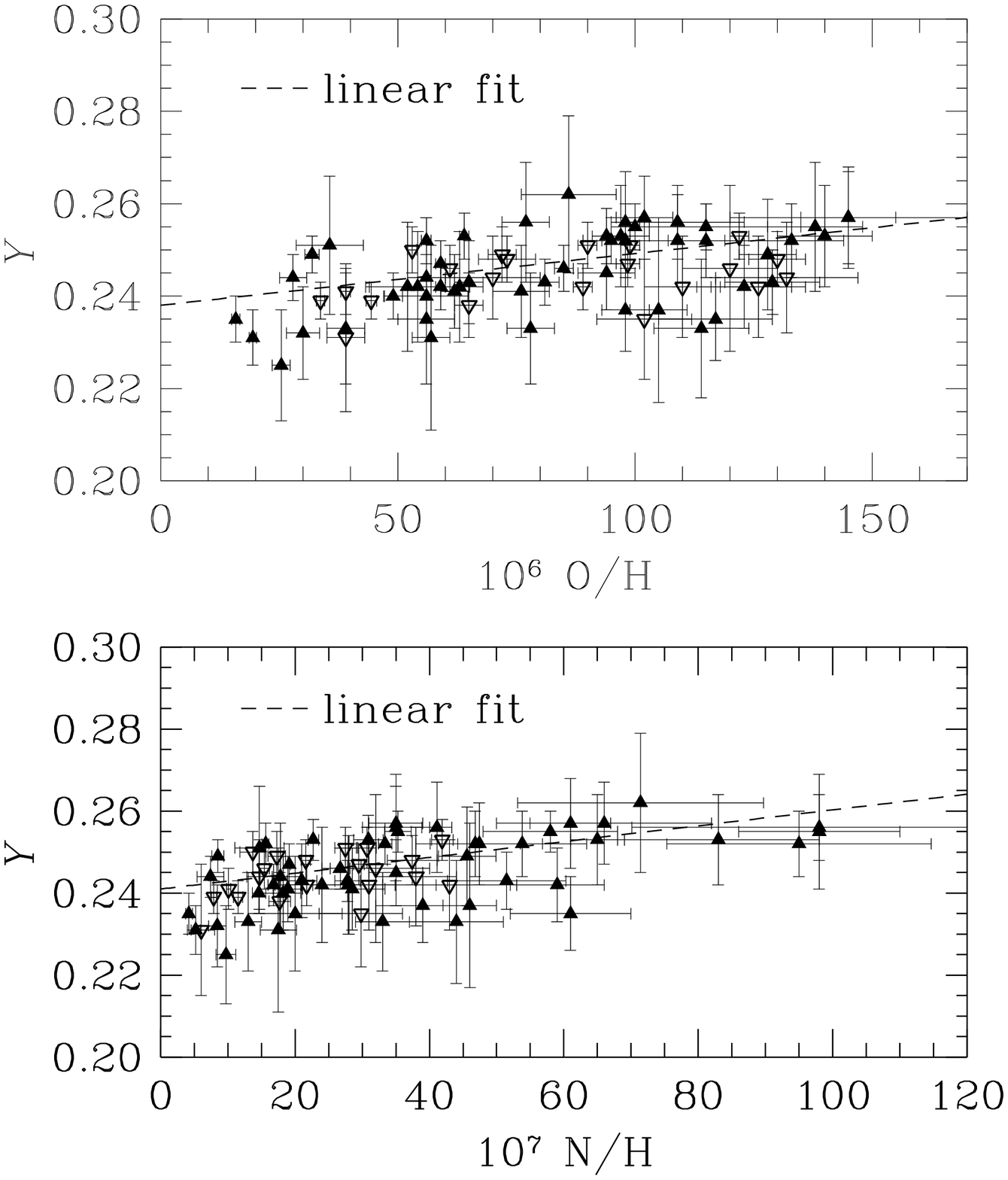}
\end{figure}  

\newpage

\begin{figure}[htb]
\epsfysize=7.5truein
\epsfbox{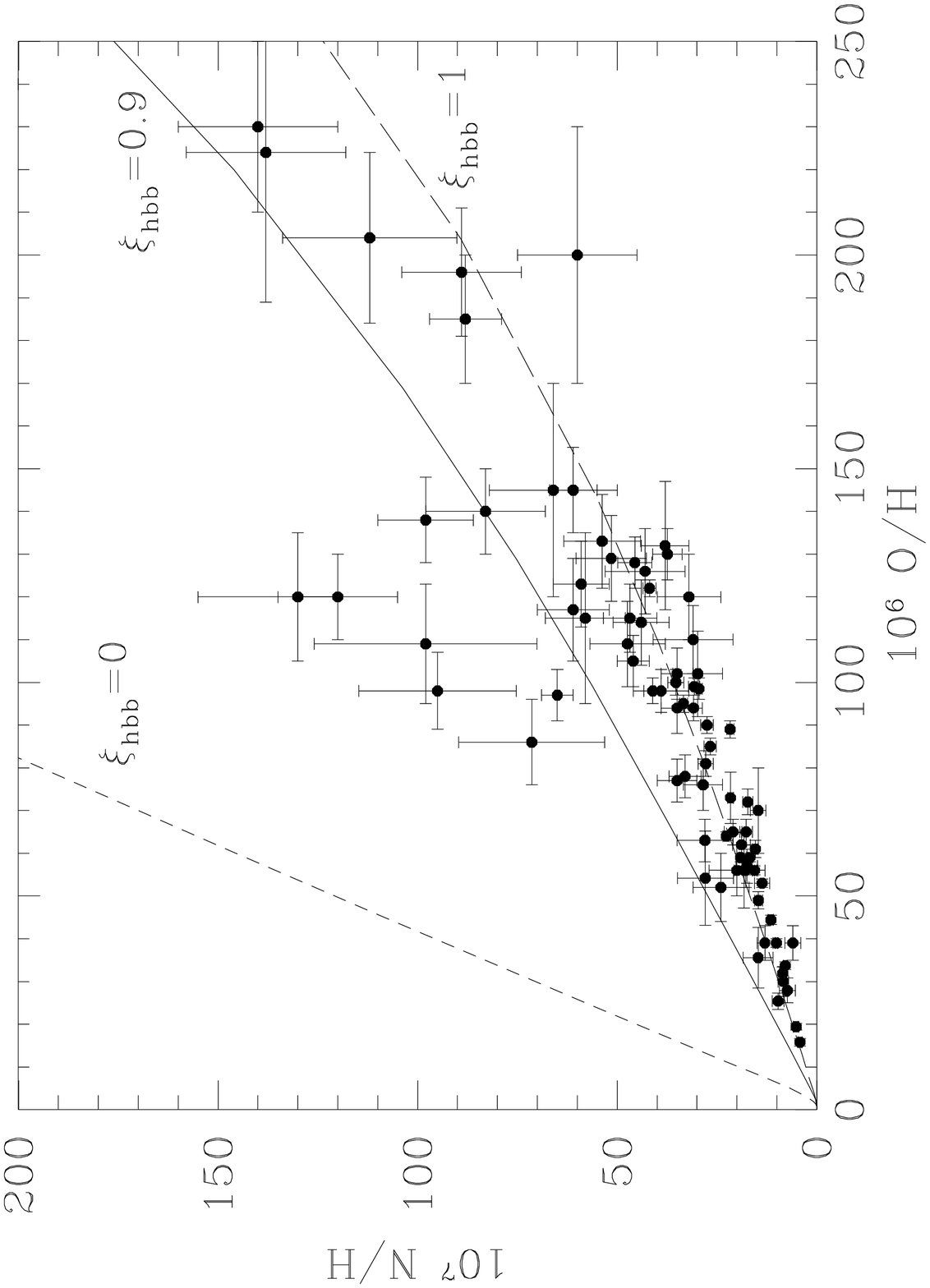}
\end{figure}

\newpage

\begin{figure}[htb]
\epsfysize=7.5truein
\epsfbox{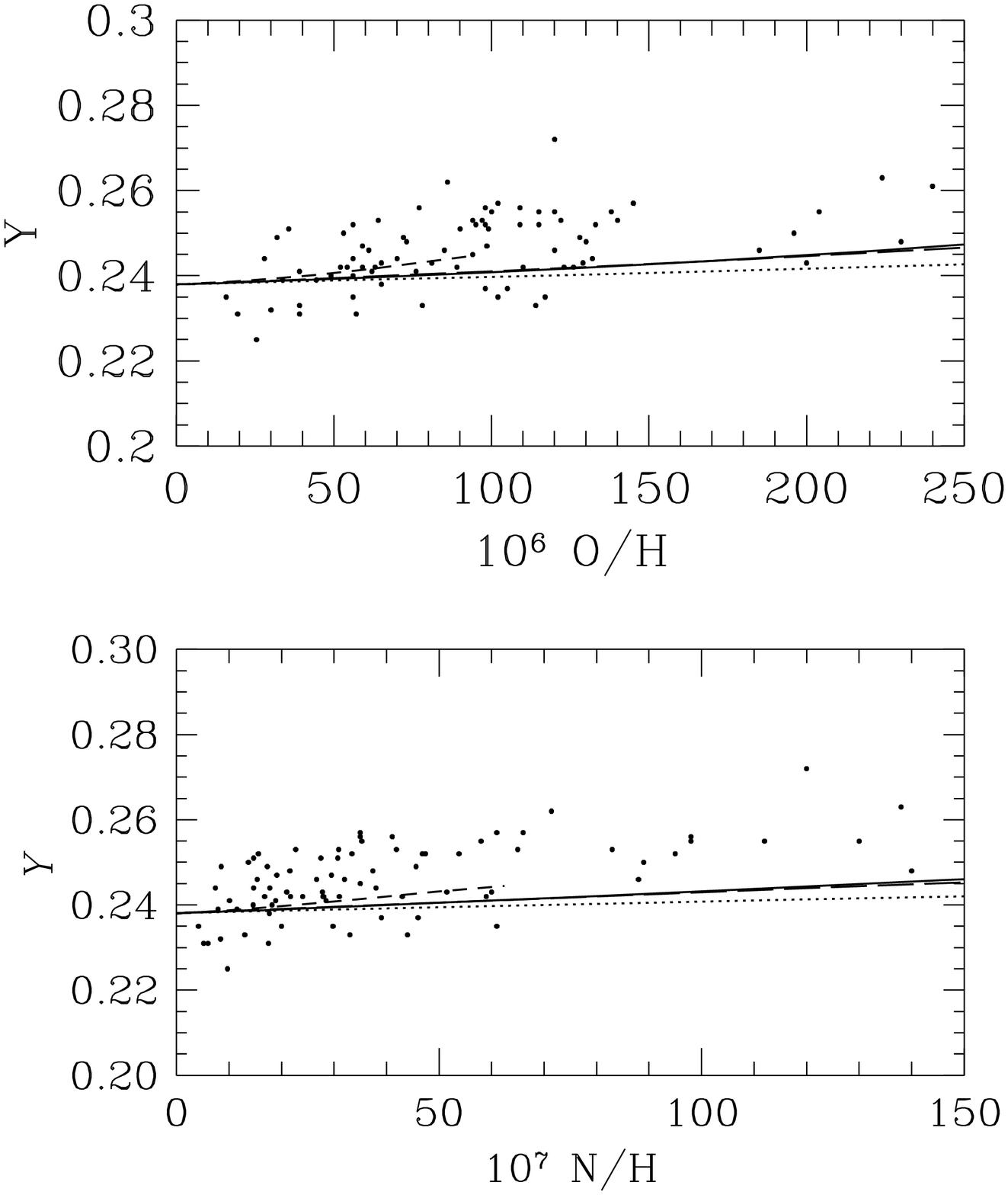}
\end{figure}

\newpage

\begin{figure}[htb]
\epsfysize=7.5truein
\epsfbox{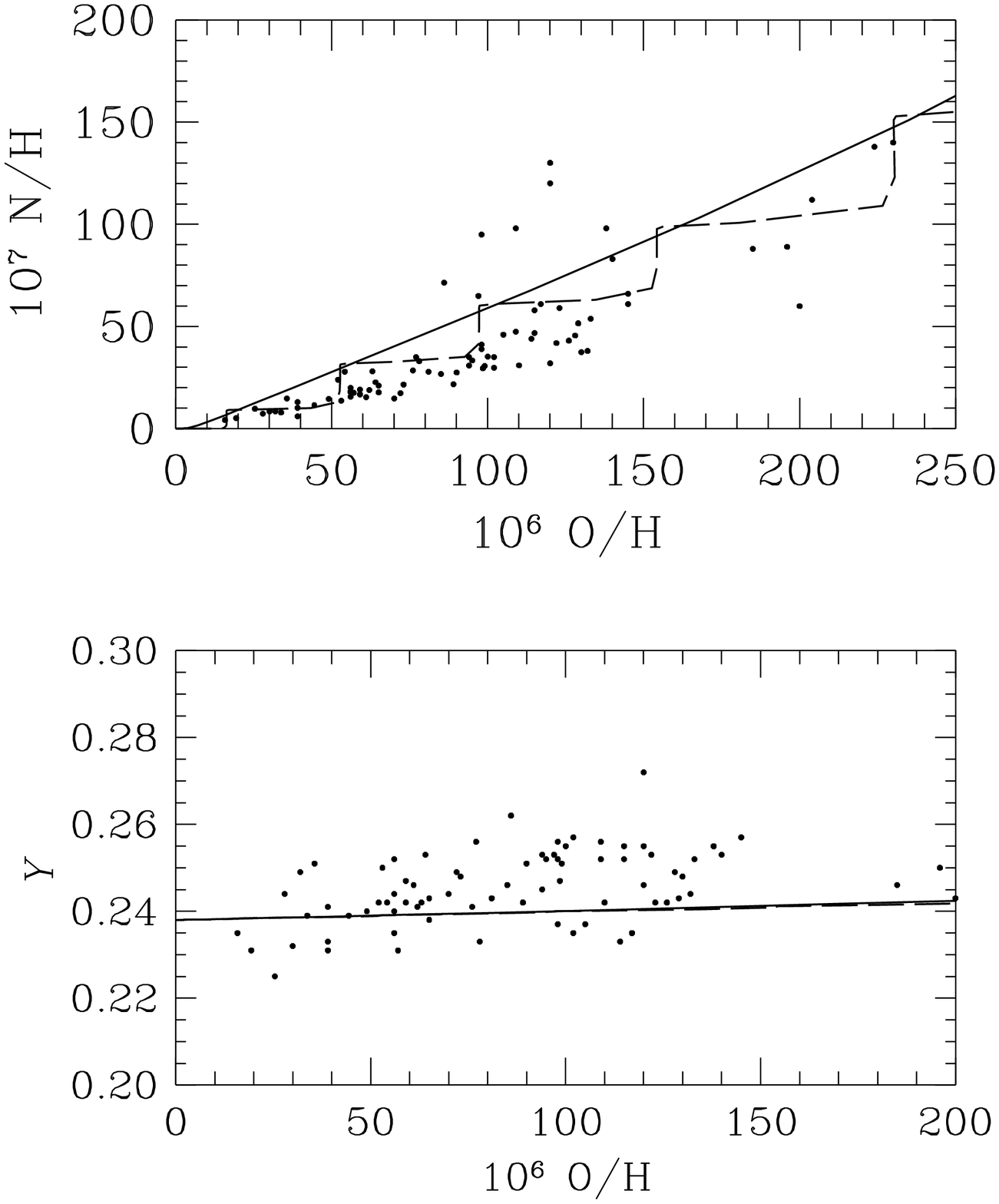}
\end{figure}

\newpage

\begin{figure}[htb]
\epsfysize=7.5truein
\epsfbox{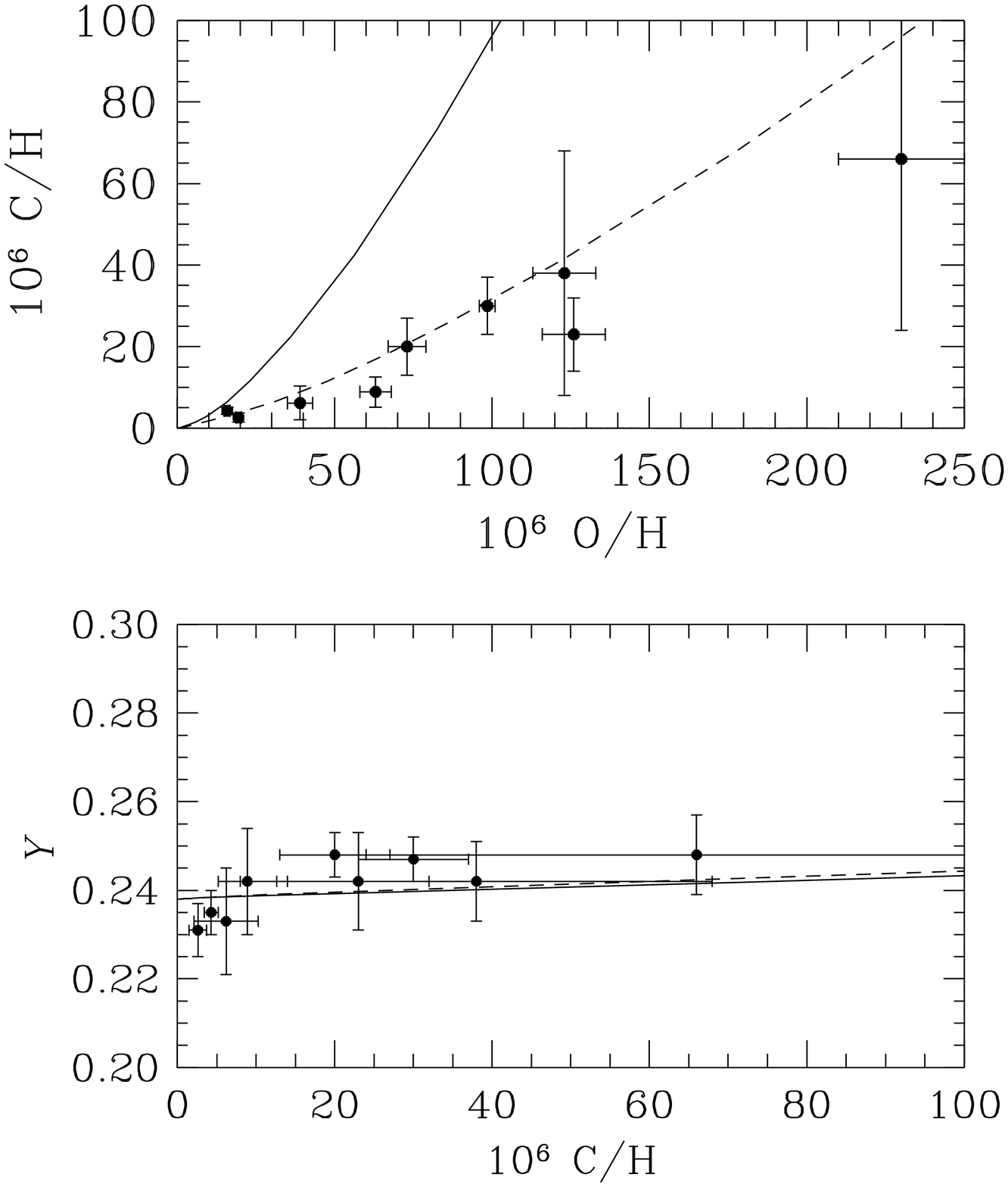}
\end{figure}
\newpage

\begin{figure}[htb]
\epsfysize=7.5truein
\epsfbox{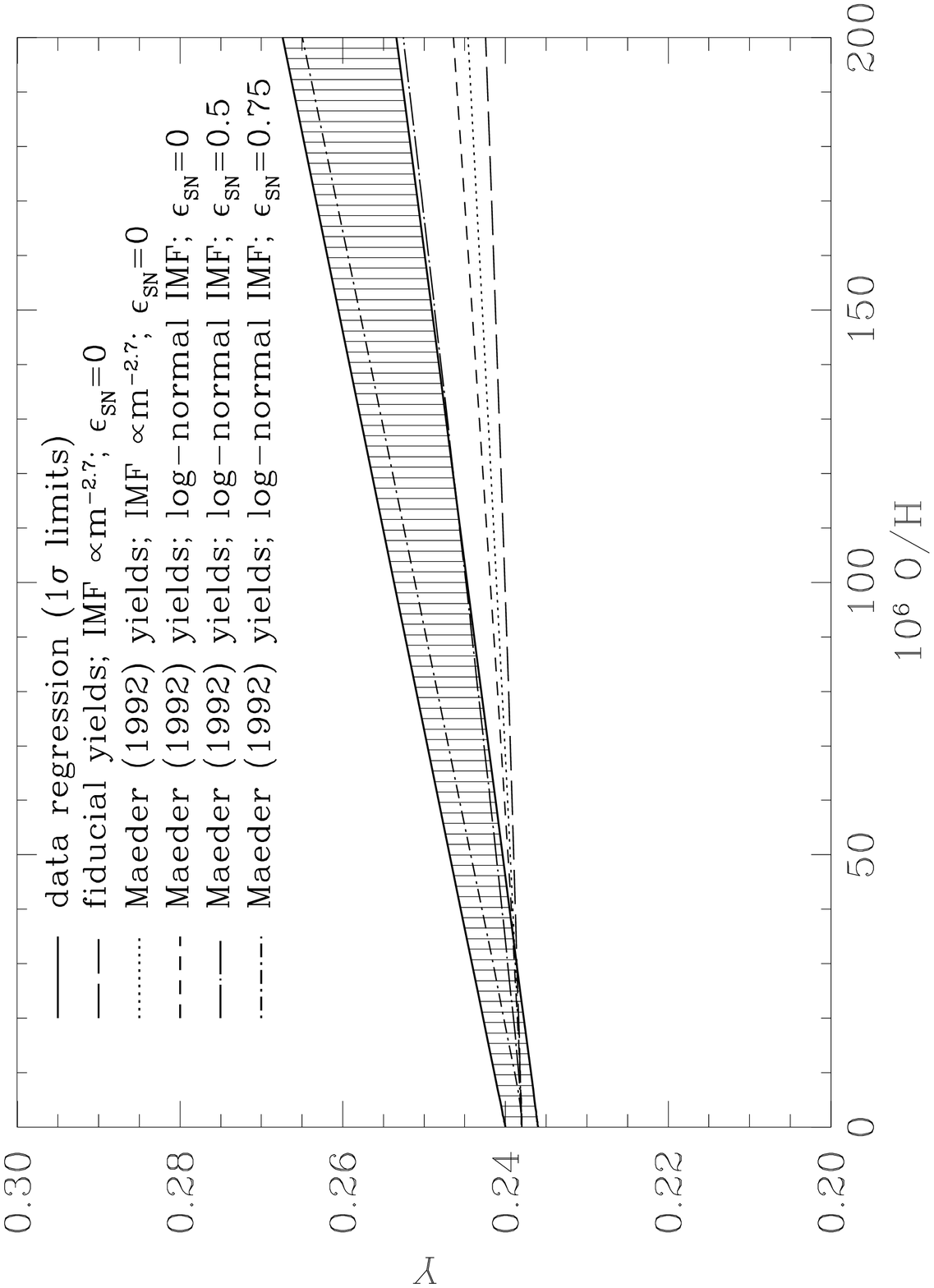}
\end{figure}

\end{document}